%% file: main.tex
\documentclass[twocolumn]{article}
\usepackage{import}
\usepackage{enumitem}
\usepackage{listings}
\usepackage{soul,color}
\usepackage{subcaption}
\usepackage{hyperref}
\hypersetup{
  hidelinks,
  colorlinks=true,
  allcolors=black,
  pdfstartview=Fit,
  breaklinks=true
}
\usepackage{multicol}
\usepackage{array, booktabs}
\usepackage{cite}
\usepackage{amsmath,amssymb,amsfonts}
\usepackage{algorithmic}
\usepackage{graphicx}
\usepackage{textcomp}
\usepackage{xcolor}
\usepackage{flushend}
\usepackage{tcolorbox}
\PassOptionsToPackage{table}{xcolor}
\def\BibTeX{{\rm B\kern-.05em{\sc i\kern-.025em b}\kern-.08em
    T\kern-.1667em\lower.7ex\hbox{E}\kern-.125emX}}
\begin{document}

\title{
To Be Forgotten or To Be Fair: Unveiling Fairness Implications of Machine Unlearning Methods
}

\author{
Dawen Zhang, Shidong Pan, Thong Hoang, Zhenchang Xing, \\Mark Staples, Xiwei Xu, Lina Yao, Qinghua Lu, Liming Zhu
\\
CSIRO's Data61
}

\date{}

\twocolumn[
  \begin{@twocolumnfalse}
  \maketitle
    \begin{abstract}

The right to be forgotten (RTBF) is motivated by the desire of people not to be perpetually disadvantaged by their past deeds. For this, data deletion needs to be deep and permanent, and should be removed from machine learning models. Researchers have proposed machine unlearning algorithms which aim to erase specific data from trained models more efficiently. However, these methods modify how data is fed into the model and how training is done, which may subsequently compromise AI ethics from the fairness perspective. To help software engineers make responsible decisions when adopting these unlearning methods, we present the first study on machine unlearning methods to reveal their fairness implications. We designed and conducted experiments on two typical machine unlearning methods (SISA and AmnesiacML) along with a retraining method (ORTR) as baseline using three fairness datasets under three different deletion strategies. Experimental results show that under non-uniform data deletion, SISA leads to better fairness compared with ORTR and AmnesiacML, while initial training and uniform data deletion do not necessarily affect the fairness of all three methods. These findings have exposed an important research problem in software engineering, and can help practitioners better understand the potential trade-offs on fairness when considering solutions for RTBF.
    \end{abstract}
  \end{@twocolumnfalse}
]

\subimport{./introduction/}{introduction}

\subimport{./background/}{background}

\subimport{./methodology/}{methodology}

\subimport{./experiments/}{experiments}

\subimport{./discussion/}{discussion}

\subimport{./threat-to-validity/}{threat-to-validity}

\subimport{./related-work/}{related-work}

\subimport{./conclusion/}{conclusion}

\input{main.bbl}

\end{document}

%% file: introduction/introduction.tex
\section{Introduction}
\label{sec:intro}

Machine learning (ML) systems
play an important role in high-stake domains. For example, ML is used to identify human faces in images and videos~\cite{singhal2022survey}, recommend products to customers~\cite{li2011design}, and recognize criminals accurately~\cite{rudin2018optimized}. ML has been called software 2.0 because its behaviours are not written explicitly by programmers, but instead are learned from large datasets~\cite{ratner2019role}.

When ML software
learns about individuals, it uses
datasets collected about them. This data contains a broad range of information that may be used to identify individuals, such as personal emails, credit card numbers, and employee records. Governments or data
subjects may sometimes ask ML service providers to remove sensitive information from their datasets for security or privacy purposes or for regulatory requirements. For example, Clearview AI\footnote{https://www.buzzfeednews.com/article/richardnieva/clearview-ordered-to-delete-in-france}, a facial recognition company owning more than 20 billion images, was requested by France’s Commission Nationale Informatique et Libertés to delete data due to a data protection law. In 2014, the Court of Justice of the European Union ordered Google, a multinational technology company, to remove links to sensitive personal data from its internet search results\footnote{https://www.reuters.com/article/us-eu-alphabet-privacy-idUSKBN1W90R5}. Later on, Europol\footnote{https://www.bleepingcomputer.com/news/security/europol-ordered-to-erase-data-on-those-not-linked-to-crime/}, the European Union Agency for Law Enforcement Cooperation, was asked to delete individuals' data having no criminal activity. Such types of demands are expected to grow in the future as regulation and privacy awareness increases.

The \textit{``right to be forgotten''} (RTBF) is covered 
in legislation in different regions, such as the General Data Protection Regulation (GDPR) in the European Union~\cite{gdpr}, the California Consumer Privacy Act (CCPA) in the United States~\cite{ccpa}, and the Personal Information Protection and Electronic Documents Act (PIPEDA) in Canada~\cite{PIPEDA}. These have given the data subject, i.e., service users, the right to request the deletion of their personal data and somehow get rid of their past~\cite{rtbf}. When ML service providers
receive such requests, they have to remove the personal data from the training set as well as update ML models to satisfy legislative purposes. Moreover, the data deletion is supposed to be deep and permanent due to the prime purpose of this right, exposing a key research challenge in various ML
applications~\cite{responsible-data-management}. 

Researchers have proposed \textit{machine unlearning} approaches to enable the RTBF to be efficiently implemented when constructing ML models. Specifically, machine unlearning is the problem of making a trained ML model forget the impact of
one or multiple data points in the training data. As ML models 
capture the knowledge learned from data, 
it is necessary to
erase what they have learned from the deleted data to fulfill the RTBF requirements. A na\"ive strategy is to retrain ML models from scratch by excluding the deleted data from the training data. However, this process may incur significant computational costs and may be practically infeasible~\cite{thompson2020computational}. Machine unlearning aims to avoid the large computational cost of fully retraining ML models from scratch and attempts to update ML models to enable the RTBF.

In recent years, machine unlearning has been extensively investigated to
address these problems~\cite{towards-unlearning, baumhauer2022machine, sisa, golatkar2020forgetting, lifelong, amnesiac}. There are two main types of machine unlearning approaches: \textit{exact machine unlearning}, and \textit{approximate machine unlearning}. While the exact machine unlearning approach ensures that the data deletion has no impact on the updated ML model by totally excluding it from the training set, and the approximate machine unlearning approach attempts to update the trained ML model weights to remove the deleted data's contribution from the trained ML model. 

Current machine unlearning research focuses on efficiency and the RTBF satisfaction, but overlooks many other critical AI properties, such as AI fairness.
\textit{AI fairness} is a non-functional property of ML software. It concerns algorithmic bias in ML models and whether they are biased toward any protected attribute classes, such as race, gender, or familial status. There is a rich literature about AI fairness~\cite{fairness-re, RE-AI, fairness-survey, fairness-testing, zhang2021ignorance, biswas2020machine, dwork2012fairness, chakraborty2020fairway}. For example, Biswas and Rajan~\cite{biswas2020machine} conducted an empirical study, employing 40 models collected from Kaggle, to evaluate the fairness of ML models. The results help AI practitioners to accelerate fairness in building ML software applications. Zhang and Harman~\cite{zhang2021ignorance} later presented another empirical study on the influence of feature size and training data size on the fairness of ML models. It suggests that when the feature size is insufficient, the ML models trained on a large training dataset have more unfairness than those trained on a small training dataset. This work also assists us to ensure ML models' fairness in practice. 

To the best of our knowledge, there is no prior work studying the fairness implications of machine unlearning methods.
However, ignoring fairness in the construction process of machine unlearning systems will adversely affect the benefit of people in  protected attribute groups such as race, gender, or familial status. For this reason, ML
systems built based on these machine unlearning methods, may violate anti-discrimination legislation, such as the Civil Rights Act~\cite{murakawa2014first}. In this paper, we conduct an empirical study to evaluate the fairness of machine unlearning models to help AI practitioners understand how to build the fairness ML systems satisfying the RTBF requirements. We aim to answer the following research questions.  

\noindent \textbf{RQ1: (Initial training)} What are the impacts of machine unlearning methods on fairness before the ``\textit{right to be forgotten}'' requests arrive?

\noindent \textbf{RQ2: (Uniform distribution)} What are the impacts of machine unlearning methods on fairness when the deleted data has uniform distribution?

\noindent \textbf{RQ3: (Non-uniform distribution)} What are the impacts of machine unlearning methods on fairness when the deleted data has non-uniform distribution?

To conduct the empirical study, we employ two popular machine unlearning methods, i.e., SISA and AmnesiacML on three AI fairness datasets.
\textbf{SISA} (Sharded, Isolated, Sliced, and Aggregated)~\cite{sisa} and \textbf{AmnesiacML}~\cite{amnesiac} are an exact machine unlearning method and an approximate machine unlearning method, respectively.
The three datasets, such as Adult, Bank, and COMPAS, have been widely used to evaluate the fairness of machine learning systems on different tasks, i.e., income prediction, customer churn prediction, and criminal detection. We use four different evaluation metrics, i.e., disparate impact, statistical parity difference, average odds difference, and equal opportunity difference, to measure the fairness of machine unlearning methods. We then analyze the results to answer the research questions.

The main contributions of our paper are as follows:
\begin{itemize}
    \item We designed and conducted an empirical study to evaluate the impacts of machine unlearning on fairness. Specifically, we employed two well-recognized machine unlearning methods on three AI fairness datasets and adopted four evaluation metrics to measure the fairness on machine unlearning systems. 
    
    \item Our results show that adopting machine unlearning methods does not necessarily affect the fairness during initial training. When the data deletion is uniform, the fairness of the resulting model is hardly affected. When the data deletion is non-uniform, SISA leads to better fairness than other methods. Through these findings, we shed light on fairness implications of machine unlearning, and provide knowledge for software engineers about the potential trade-offs when selecting solutions for RTBF.
    
\end{itemize}

%% file: background/background.tex
\section{Background}
\label{sec:background}

This section provides the background knowledge, including machine unlearning methods and AI fairness metrics.

\subsection{Machine Unlearning Methods}
\label{sec:unlearning}

The classification problem is a type of task that many machine learning systems aim to solve and in which machine unlearning can be leveraged.
Given a dataset of input-output pairs $\mathcal{D}= (x, y) \in \mathcal{X} \times \mathcal{Y}$, we aim to construct a prediction function $\mathcal{F}_{\mathcal{D}}: \mathcal{X} \rightarrow \mathcal{Y}$ that maps these inputs to outputs. The prediction function $\mathcal{F}_{\mathcal{D}}$ is often learned by minimizing the following objective function:
\begin{equation}
\underset{\mathcal{F}_{\mathcal{D}}}{min} \sum_{i}\mathcal{L}(\mathcal{F}_{\mathcal{D}}(x_i), y_i) + \lambda\Omega(\mathcal{F}_{\mathcal{D}})
\end{equation}
where $\mathcal{L}(.)$, $\Omega(\mathcal{F}_{\mathcal{D}})$, and $\lambda$ are the empirical loss function, the regularization function, and the trade-off value, respectively.
Let $\mathcal{D}_{r}$ and $\mathcal{D}_{u}$ represent the retained dataset and the deleted dataset respectively. $\mathcal{D}_{r}$ and $\mathcal{D}_{u}$ are mutually exclusive, i.e., $\mathcal{D}_{r} \cap \mathcal{D}_{u} = \O \mathcal{}$ and $\mathcal{D}_{r} \cup \mathcal{D}_{u} = \mathcal{D}$. When the \textit{``right to be forgotten''} (RTBF) requests arrive, a machine unlearning system needs to remove $\mathcal{D}_{u}$ from $\mathcal{D}$ and update the prediction function $\mathcal{F}_{\mathcal{D}}$. Machine unlearning attempts to achieve a model $\mathcal{F}_{\mathcal{D}_{r}}$, only trained from the retained dataset $\mathcal{D}_{r}$, without incurring a significant computational cost. Hence, the model $\mathcal{F}_{\mathcal{D}_{r}}$ is often used to evaluate the performance of machine unlearning methods. 

There are mainly two types of machine unlearning approaches, such as exact machine unlearning and approximate machine unlearning.
We present a typical method for each machine unlearning approach. Specifically, SISA and AmnesiacML are selected to represent the exact machine unlearning approach and the approximate machine unlearning approach, respectively. These methods, adopted for deep learning models, are efficient and effective in dealing with RTBF requests. We will briefly describe them in the following subsections.

\begin{figure}[t!]
  \centering
  \includegraphics[width=1.0\linewidth]{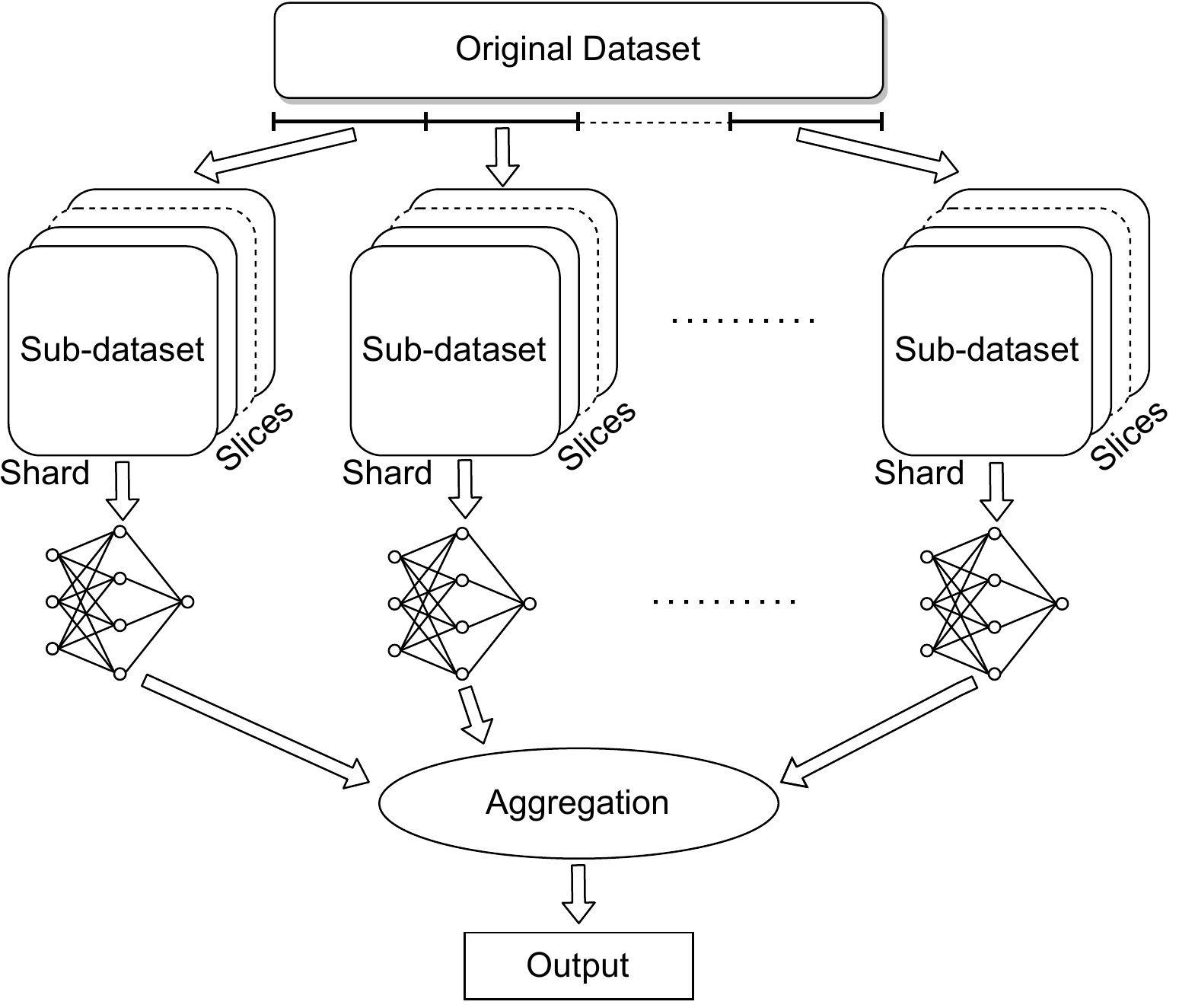}
  \caption{An overview framework of SISA. The dataset is first sharded into multiple shards. Each shard is further sliced into multiple slices. Each shard is put into a deep learning model trained by gradually increasing the number of slices. The output of the DL models is combined using a voting-based aggregation.}
  \label{fig:sisa}
\end{figure}

\subsubsection{SISA~\cite{sisa}} This is an exact machine unlearning method aiming to reduce the computational cost of the retraining process by employing a data partitioning technique. Figure~\ref{fig:sisa} briefly describes an overview framework of SISA. In the beginning, the original data $\mathcal{D}$ is split into $\mathcal{S}$ shards, such as $\cap_{i \in |\mathcal{S}|} D_i = \O$ and $\cup_{i \in |\mathcal{S}|} D_i = \mathcal{D}$.
Each shard $D_i \in \mathcal{D}$ is then further split into $K$ slices, i.e.,  $\cap_{k \in |K|} D_{ik} = \O$ and $\cup_{k \in |K|} D_{ik} = D_i$. 
A deep learning (DL) model is constructed on each shard. The DL model is updated by gradually increasing the number of slices. Note that all the parameters of the DL model are kept in storage. After finishing the training process, SISA contains multiple DL models. Finally, the output results are collected by employing a voting mechanism on a list of outputs of DL models. When RTBF requests arrive, SISA automatically locates the shards and the slices containing the deleted data $\mathcal{D}_u$. SISA then retrains the DL models of these shards from the particular cached stage, i.e., before the slices of the deleted data were put into the DL models.

\begin{figure}[htbp]
  \centering
  \begin{subfigure}[b]{0.235\textwidth}
         \centering
  \includegraphics[width=0.95\linewidth]{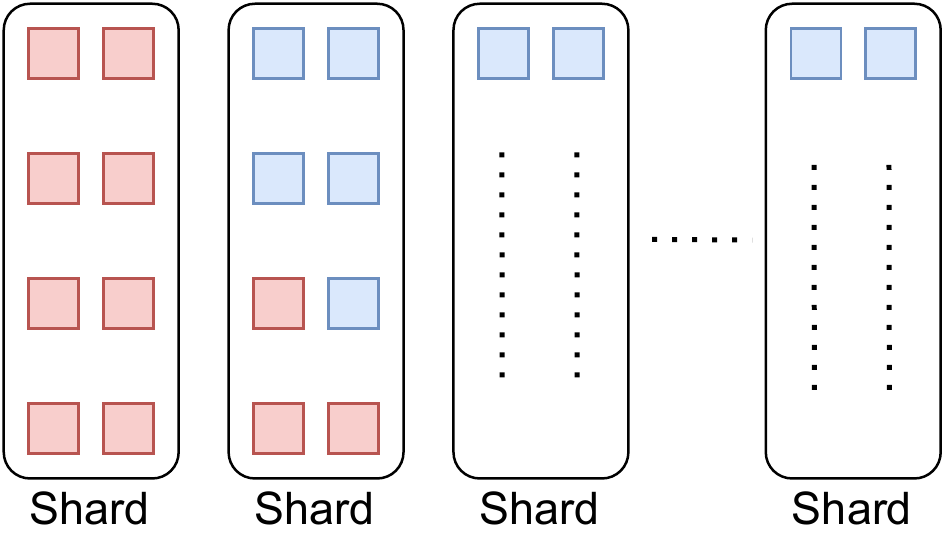}
  \caption{Instances with a higher deletion probability, illustrated as red squares, are allocated to the same shards so that when RTBF requests arrive, fewer shards are required to be retrained compared with the naive way of randomly allocating the instances.}
  \label{fig:sisa-sharding}
  \end{subfigure}
  \hspace{0.5mm}
  \begin{subfigure}[b]{0.235\textwidth}
  \centering
  \includegraphics[width=\linewidth]{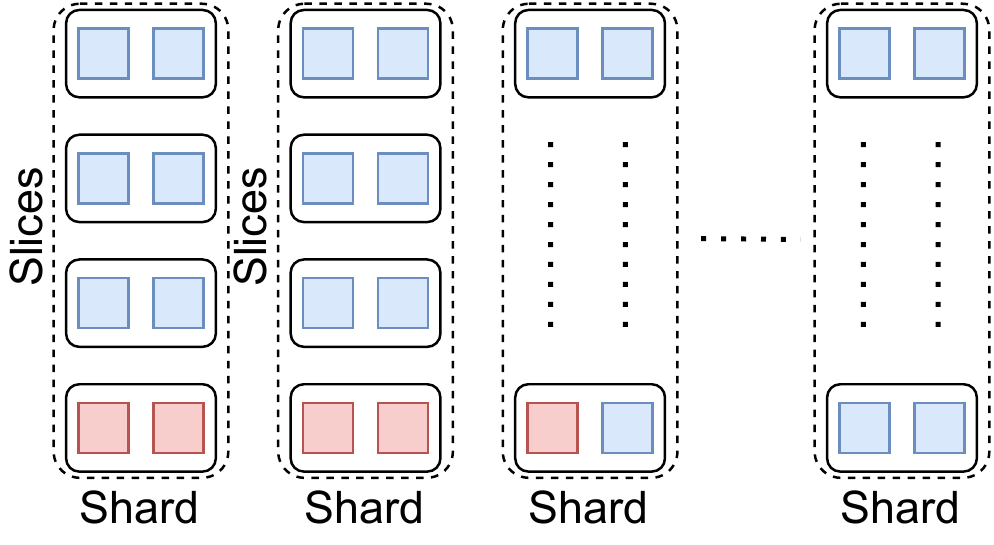}
  \caption{Instances with high likelihood, illustrated as red squares, are placed in the last slices. When we receive the RTBF requests, the DL model of each shard is retrained from a checkpoint before those slices containing the deleted data are merged. 
  }
  
  \label{fig:sisa-slicing}
    \end{subfigure}
   \caption{SISA's strategies aim to reduce the computational cost of the retraining process.}
\end{figure}

The priori probability is the probability of an event happening when we have a limited number of possible outcomes that equally occur~\cite{leung2007naive}. Machine unlearning methods can easily improve their performance when we know the priori probability of data deletion from different groups. For example, wealthy families prefer to keep their privacy for safety purposes, so they tend to send RTBF requests compared to other people~\cite{upton2001strategic}. Another example is that people with a higher educational background are more likely to remove their personal information from public~\cite{eurobarometer}.

There are two strategies for SISA to leverage the priori probability to speed up the training process, hence reducing the computational cost. The first strategy is to allocate the instances with a higher deletion probability into the same shards. This means the retraining process would happen on fewer shards compared with randomly allocating the instances. The second strategy is to allocate the instances with a higher deletion probability to the last slices. In this case, the retraining process would happen on fewer slices compared with randomly allocating the instances. Figure~\ref{fig:sisa-sharding} and Figure~\ref{fig:sisa-slicing} briefly describe the first and second strategies, respectively.

SISA is efficient and effective in dealing with machine unlearning problems. The method has inspired many later works~\cite{recommendation, coded, graph-eraser}. Its source code is placed at \url{https://github.com/cleverhans-lab/machine-unlearning}.

\subsubsection{AmnesiacML~\cite{amnesiac}} This is a method of approximate machine unlearning approach. AmnesiacML makes use of the characteristics of batch training in neural networks. During the training process, the updated parameters of a DL model for each batch are recorded and kept in storage. The training process is expressed as follows:

\begin{equation}
\theta_{M} = \theta_{\mathrm{initial}} + \sum_{e=1}^{E}\sum_{b=1}^{B}\Delta_{\theta_{e,b}}
\label{amnesiac-formula}
\end{equation}
where $\theta_{\mathrm{initial}}$ is the initial parameters of the DL model, $E$ and $B$ represent the total number of epochs and the total number of batches in each epoch, respectively. The updated parameters are stored as $\{ \gamma_b \mid \gamma_b = \sum_{e=1}^E\Delta_{\theta_{e,b}} , 1 \leq b \leq B\}$.

When we receive the RTBF requests, AmnesiacML automatically locates the batches containing the instances that need to be deleted. After that, the DL model's parameters are rolled back to remove the impact of the deleted data on the trained DL model as follows: 

\begin{equation}
\theta_{M'} = \theta_{M} - \sum_{\hat{b}=1}^{\hat{B}}\gamma_{\hat{b}}
\label{amnesiac-update}
\end{equation}

A strategy for AmnesiacML is easily adopted when we comprehend the priori probability of deleted data from different groups. For example, instances with a higher priori probability of being removed can be placed into the same batches. Hence, the process of updating parameters in the DL model will require a less computational cost.

Similar to SISA, AmnesiacML shows its efficiency and effectiveness in machine unlearning problems. However, it does not ensure the impact of deleted data being completely forgotten in the updated DL model. The open-source repository of AmnesiacML can be found at \url{https://github.com/lmgraves/AmnesiacML}

\subsection{AI Fairness Metrics}
\label{sec:ai_metrics}

The goal of AI fairness is to correct machine learning (ML) models with the assumption that models should not be biased between any protected classes, i.e., race, sex, familial status, etc. Each protected class partitions a population into different groups, such as the privileged group and the unprivileged group. 
In this section, we employ four different fairness metrics, such as disparate impact, statistical parity difference, average odds difference, and equal opportunity difference, to evaluate the impact of machine unlearning methods on fairness. These metrics are widely adopted in measuring the fairness of ML systems~\cite{fairness-re, RE-AI, fairness-survey, fairness-testing, zhang2021ignorance, biswas2020machine, dwork2012fairness, chakraborty2020fairway}.

Let $x_s \in \{0, 1\}$ indicates the binary label of a protected class ($x_s = 1$ for the privileged group). Let $\hat{y} \in \{0, 1\}$ be the predicted outcome of a ML classification model ($\hat{y}=1$ for the favourable decision). Let $y \in \{0, 1\}$ be the binary classification label ($y=1$ is favourable). We present the four fairness evaluation metrics as follows.

\noindent \textbf{Disparate impact (DI)}~\cite{di} measures the ratio of the favourable outcome of the unprivileged group ($x_s=0$) against the privileged group ($x_s=1$). 
\begin{equation}
    \frac{P[\hat{y} = 1 \mid x_s = 0]}{P[\hat{y} = 1 \mid x_s = 1]}
\label{di-formula}
\end{equation}

\noindent \textbf{Statistical parity difference (SPD)}~\cite{spd} is the difference of the favourable outcome of the unprivileged group ($x_s=0$) against the privileged group ($x_s=1$). 
\begin{equation}
    P[\hat{y} = 1 \mid x_s = 0] - P[\hat{y} = 1 \mid x_s = 1]
\label{spd-formula}
\end{equation}

\noindent\textbf{Average odds difference (AOD)}~\cite{equal-op} calculates the average of difference in true positive rate and false positive rate between unprivileged and privileged groups. 
\begin{equation}
\begin{split}
    \frac{1}{2} (\lvert P[\hat{y} = 1|x_s = 0, y=1] - P[\hat{y} = 1|x_s = 1, y=1] \rvert \\
    +\lvert P[\hat{y} = 1|x_s = 0, y=0] - P[\hat{y} = 1|x_s = 1, y=0]\rvert)
\end{split}
\label{aod-formula}
\end{equation}

\noindent \textbf{Equal opportunity difference (EOD)}~\cite{equal-op} evaluates the difference in true positive rate between unprivileged group and privileged groups.
\begin{equation}
    P[\hat{y} = 1|x_s = 0, y=1] - P[\hat{y} = 1|x_s = 1, y=1]
\label{eod-formula}
\end{equation}

All fairness metrics are range from -1 to 1. Among them, DI achieves the greatest fairness of the classification model when it equals 1. The remaining fairness metrics, i.e., SPD, AOD, and EOD, attain the greatest fairness when their values are 0.

%% file: methodology/methodology.tex
\section{Methodology}
\label{sec:method}

This section first describes our experimental design and setup. Then we briefly present the datasets, the data deletion strategies, and our evaluation metrics. 

\begin{figure*}[t!]
  \centering
  \includegraphics[scale=0.8]{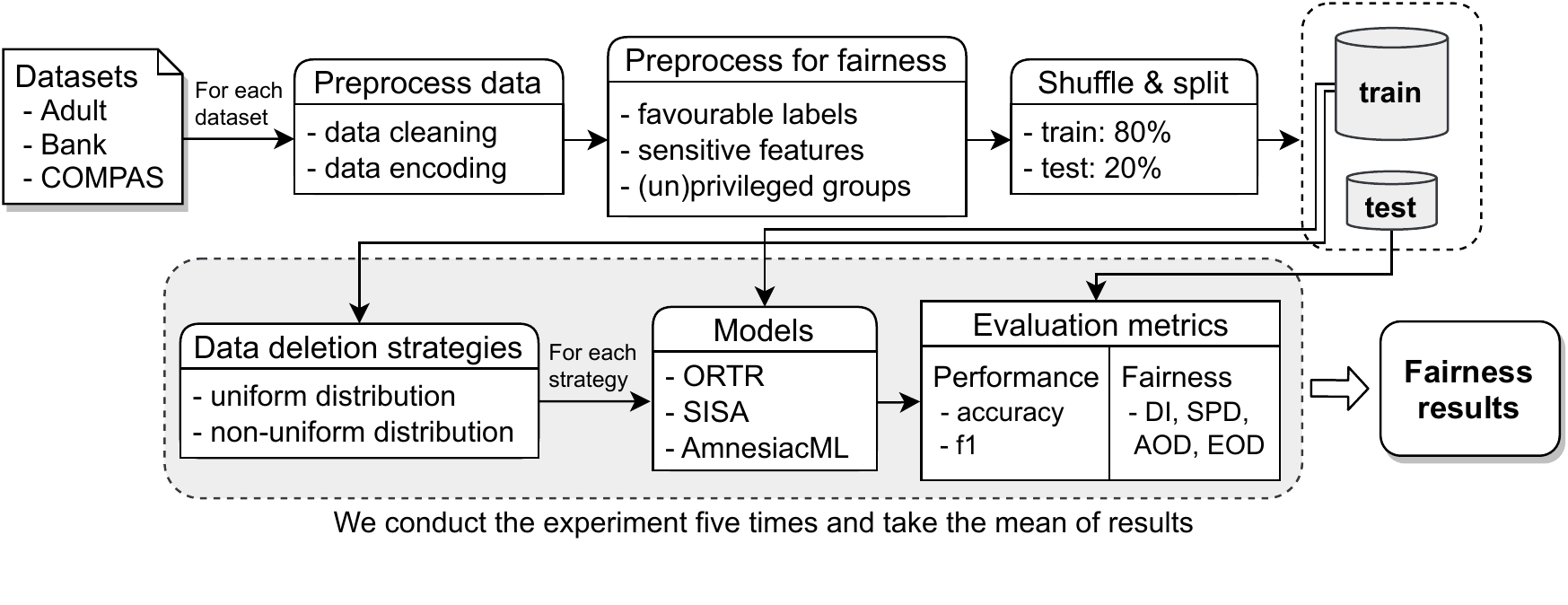}
  \caption{Experimentation to evaluate the performance and fairness of machine unlearning methods under different scenarios. 
  }.
  \label{fig:design}
\end{figure*}

\subsection{Experiment Design}
\label{sec:design}

Our empirical study starts by first collecting the benchmark fairness datasets. For each dataset, we preprocess and split it into training and testing datasets. The training dataset is then employed to train machine unlearning models. We use six evaluation metrics to measure the performance and fairness of these models. Figure~\ref{fig:design} briefly presents an overview framework of our experimental design. 

To identify the fairness datasets, we first refer to the work on fairness testing for machine learning models that employed six datasets, such as German Credit, Adult, Bank, US Executions, Fraud Detection, and Raw Car Rentals~\cite{aggarwal2019black}. Among these datasets, only German Credit, Adult, and Bank are available. We also collect the Heart Disease dataset~\cite{heart}, referring to the presence of heart disease in patients, and the COMPAS dataset~\cite{compas}, aiming to predict the probability of criminals reoffending. In total, we acquire five datasets, i.e., German Credit, Adult, Bank, Heart Disease, and COMPAS, across various domains. As machine unlearning methods are efficient and effective on large datasets~\cite{sisa, amnesiac}, we remove datasets that have fewer than 1,000 instances including German Credit and Heart Disease. Hence, there are three datasets, i.e., Adult, Bank, and COMPAS, that are employed to evaluate the impacts of machine unlearning methods in our experiments. 

We apply the same data preprocessing approach for all three datasets. Specifically, we employ the AI Fairness 360 toolkit~\cite{aif360}, which is an open-source library for fairness metrics, to clean up invalid or missing values, transform categorical values into a one-hot encoding, and convert non-numerical binary values to a binary label (e.g., \textit{male}: 1, \textit{female}: 0). We further preprocess the datasets to employ them for fairness evaluation. Specifically, we specify favourable labels or the predicted outcome of our model. We also identify sensitive features (or protected classes) for the privileged and unprivileged groups.
For example, in the Adult dataset, the prediction label is a favourable label, indicating whether a person has a high annual salary. We define \textit{sex} as a sensitive feature. We assume that a \textit{male} often has a higher annual salary than a \textit{female}; hence, the \textit{male} should be put in the privileged group while the \textit{female} should be in the unprivileged group regarding the sensitive feature \textit{sex}. 
For each dataset, we shuffle and split it into the training dataset (80\%) and the testing dataset (20\%). We then feed the training dataset into our models. 

To conduct our experiments, we employ a multilayer perceptron (MLP), a simple feedforward network~\cite{ramchoun2016multilayer}. The MLP model includes an input layer, a hidden layer, and an output layer. We train the MLP model by optimizing a cross-entropy loss function~\cite{martinez2018taming}. Two machine unlearning methods, such as SISA and AmnesiacML, are built based on the MLP model. A na\"ive approach of using original training and retraining (denoted as \textbf{ORTR}) is also built based on the MLP model as the baseline. We consider two experimental scenarios. 

\begin{itemize} [leftmargin=*]
    \item \textbf{Scenario 1:}
    Before any ``\textit{right to be forgotten}'' (RTBF) requests, what are the impacts of machine unlearning methods on fairness? In this setting, the training dataset is put into three different models, such as ORTR, SISA, and AmnesiacML (see Figure~\ref{fig:design}), to train these models. We then employ the testing dataset to evaluate the performance and fairness of these trained models. 
    \item \textbf{Scenario 2:} When the RTBF requests arrive, what are the impacts of machine unlearning methods on fairness? In this setting, we employ data deletion strategies (see Figure~\ref{fig:design}) to remove instances from the training dataset. For each data deletion strategy, we compare the performance and fairness of ORTR with two machine unlearning methods, such as SISA and AmnesiacML. 
\end{itemize}

For each dataset, we apply 5-fold cross-validation and take the mean of the results. We have conducted our experiments using an Nvidia T4 GPU and an Intel Xeon Silver 4114 CPU with 16 GB RAM and 12 GB RAM, respectively. The OS is Debian 10.10 LTS 64 bit. The machine learning framework is PyTorch v.1.12 with CUDA 11.3, and the Python language version is 3.7.

\subsection{Datasets}
\label{sec:data}

We conduct our experiments by employing three widely-used fairness datasets to evaluate the impacts of machine unlearning methods on fairness. These datasets are briefly described as follows. 

\begin{itemize} [leftmargin=*]
    \item \textbf{Adult}~\cite{adult}. This dataset is extracted from the 1994 Census Bureau database\footnote{https://www.census.gov/programs-surveys/ahs/data/1994.html}. Its task is to predict whether a person can earn over \$50,000 USD per year. The dataset includes 48,842 instances and 14 features. The sensitive features for this dataset are \textit{sex} and \textit{race}. 
    \item \textbf{Bank}~\cite{bank}. The dataset is collected from marketing campaigns of a Portuguese banking institution. Its task is to predict whether a client will subscribe to a bank term deposit. The dataset contains 45,211 instances and 17 features. We use \textit{age} as the sensitive feature for this dataset.
    \item \textbf{COMPAS}~\cite{compas}. The dataset contains recidivism records, which are used to build a prediction system to forecast the possibility of a criminal defendant reoffending. The dataset has 7,215 instances and seven features. The sensitive features are defined as \textit{sex} and \textit{race}.
\end{itemize}

All the sensitive features are selected by following the previous work~\cite{biswas2020machine, chakraborty2020fairway, zhang2021ignorance}.

\subsection{Data Deletion Strategies}
\label{sec:data_deletion}

To send the ``\textit{right to be forgotten}'' (RTBF) requests, we adopt two data deletion strategies. Each strategy has various settings presented as follows.

\noindent {\textbf{Uniform distribution.}} For this strategy, we assume that the deleted data has a uniform distribution, i.e., each instance has an equal probability of being removed from the training dataset. To select a range of proportions of the total amount of deleted data, we leverage the work of Bertram et al~\cite{rtbf5years}. Specifically, we randomly remove 1\%, 5\%, 10\%, and 20\% of the training data. 

\noindent {\textbf{Non-uniform distribution.}} For this strategy, we assume that the deleted data has a non-uniform distribution, i.e., each instance has a different probability of being removed from the training dataset. Some people have a higher probability of sending RTBF requests compared to other people. For example, people who are from wealthy families or have a high educational background prefer to keep their sensitive information private for security and privacy purposes~\cite{upton2001strategic, eurobarometer}. 
As these personal details are unavailable in our datasets, to better understand the fairness implications under different cases when the deleted data is a non-uniform distribution, we first assume that the people who request the RTBF are predominantly from privileged groups, and we assume another scenario that people exercising the RTBF are predominantly from unprivileged groups.

\subsection{Evaluation Metrics}
\label{sec:metrics}

We consider two types of evaluation metrics in our experiments, which are performance and fairness. 

\noindent \textbf{Performance measure.} Before evaluating the fairness of models, we calculate their performance in terms of accuracy and F1 score. 

\begin{itemize} [leftmargin=*]
    \item \textit{Accuracy:} The ratio of true predictions among the total number of predictions~\cite{gunawardana2009survey}.
    \item \textit{F1 score:} The harmonic mean between precision and recall~\cite{dalianis2018evaluation}.
\end{itemize}

\noindent \textbf{Fairness measure.} To measure the fairness of models, we adopt the four fairness metrics, i.e., disparate impact (DI), statistical parity difference, average odds difference, and equal opportunity difference, briefly mentioned in Section~\ref{sec:ai_metrics}. For simplicity in presenting and observing, we convert all the fairness metric values into their absolute values. As disparate impact (DI) value differs from other fairness metrics, we use $|$1 - DI$|$ to evaluate the fairness of our models. In this case, all four fairness metrics achieve the greatest fairness when their values equal 0.

%% file: experiments/experiments.tex
\section{Experiments}
\label{sec:experiment}

In this section, we provide results and insights from the experimentation, to answer our research questions.

\begin{figure}[t!]
  \centering
   \begin{subfigure}[b]{0.24\textwidth}
         \centering
         \includegraphics[width=\textwidth]{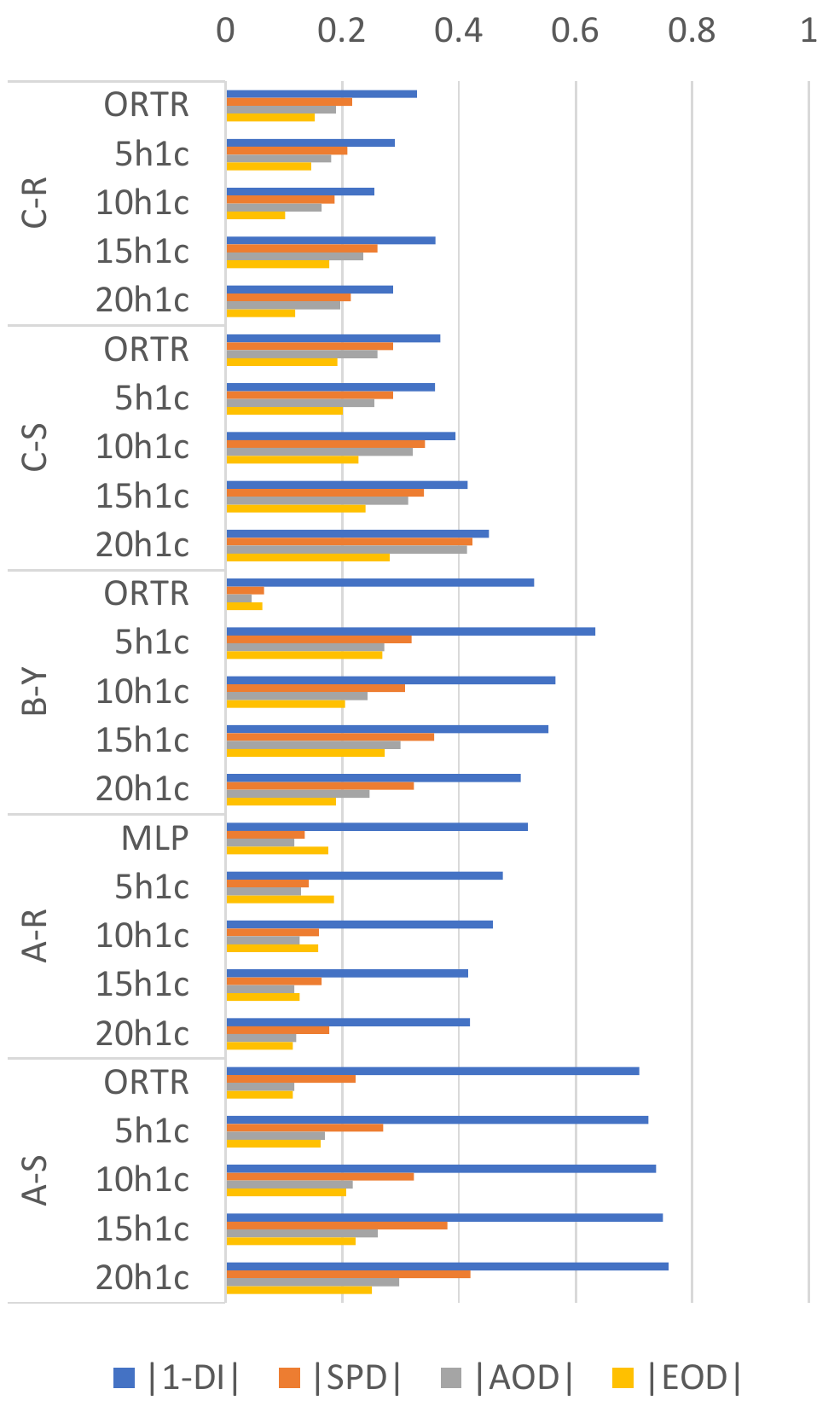}
         \caption{Fairness metrics for \\ 5/10/15/20 shards one slice
         }
         \label{fig:rq1-fairness-1slice}
 \end{subfigure}
    \begin{subfigure}[b]{0.24\textwidth}
         \centering
         \includegraphics[width=\textwidth]{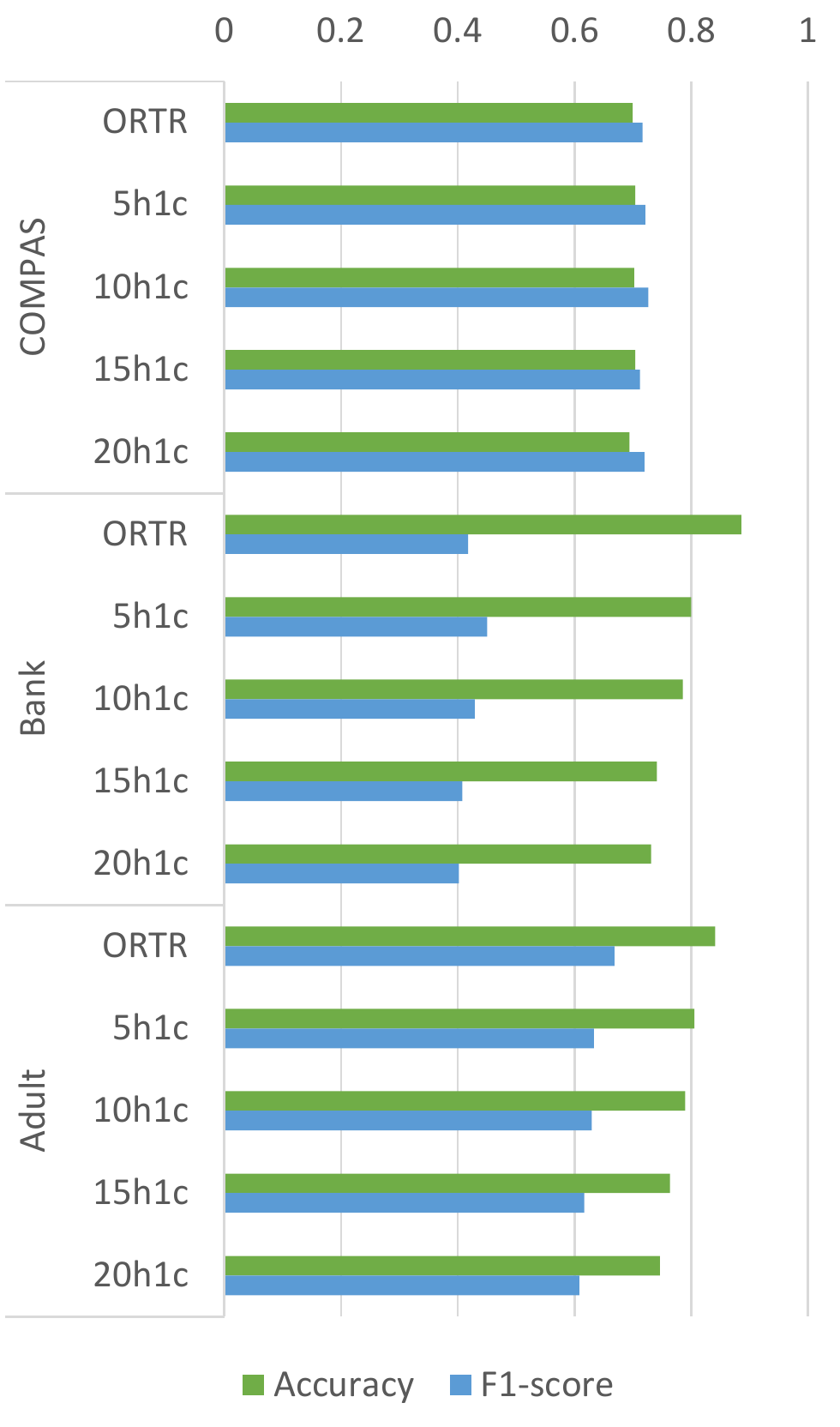}
         \caption{Performance metrics for \\ 5/10/15/20 shards one slice}
         \label{fig:rq1-performance-1slice}
 \end{subfigure}

 \begin{subfigure}[b]{0.24\textwidth}
         \centering
         \includegraphics[width=\textwidth]{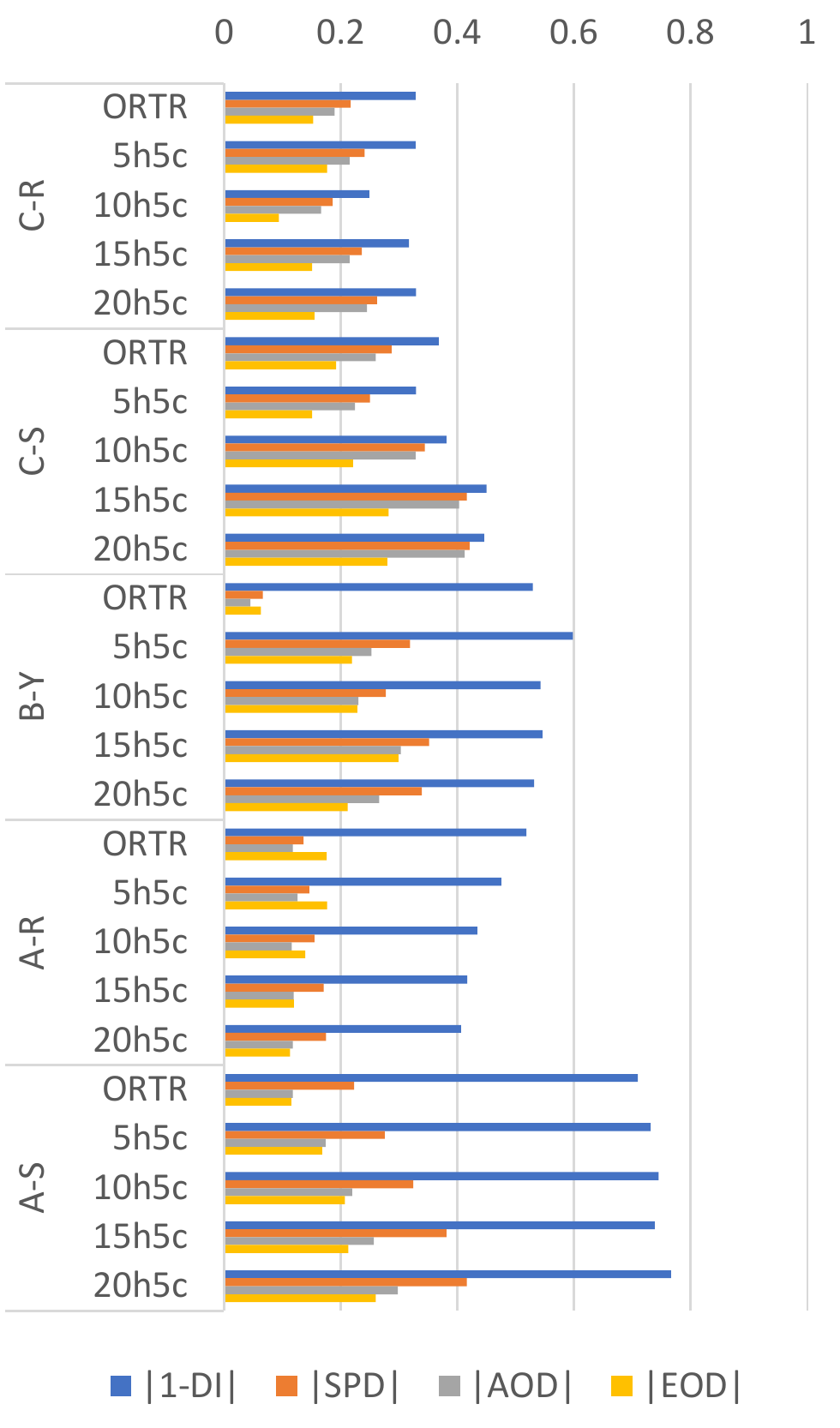}
         \caption{Fairness metrics for \\ 5/10/15/20 shards five slices}
         \label{fig:rq1-fairness-5slice}
 \end{subfigure}
    \begin{subfigure}[b]{0.24\textwidth}
         \centering
         \includegraphics[width=\textwidth]{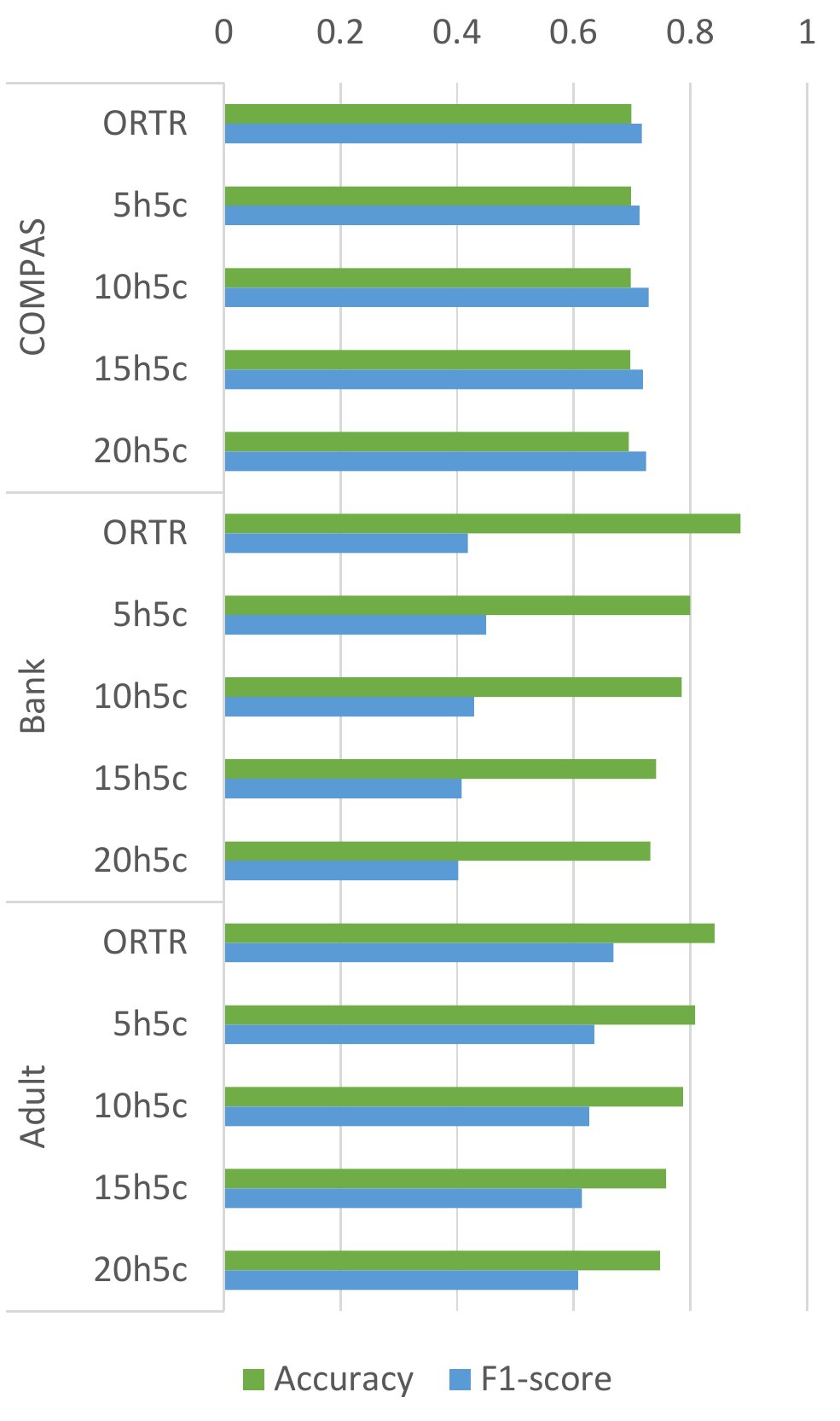}
         \caption{Performance metrics for \\ 5/10/15/20 shards five slices}
         \label{fig:rq1-performance-5slice}
 \end{subfigure}
  \caption{Fairness (the smaller, the better) and performance (the higher, the better) evaluation results of SISA with different shards (5/10/15/20h) and slices (1/5/10c)}
  \label{fig:rq1}
\end{figure}

\begin{figure*}[t!]
  \centering
  \includegraphics[width=1.0\textwidth]{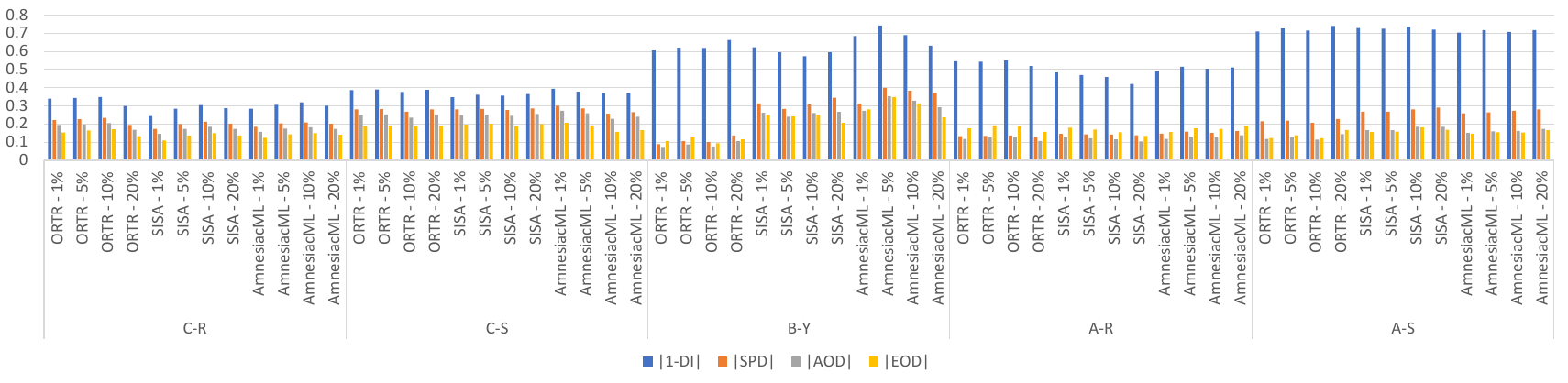}
  \caption{Fairness (the smaller, the better) evaluation results of different training methods after uniform data deletion under various deletion proportions.}
  \label{fig:rq2-bar-chart}
\end{figure*}

\begin{figure*}[t!]
  \centering
  \includegraphics[width=1.0\textwidth]{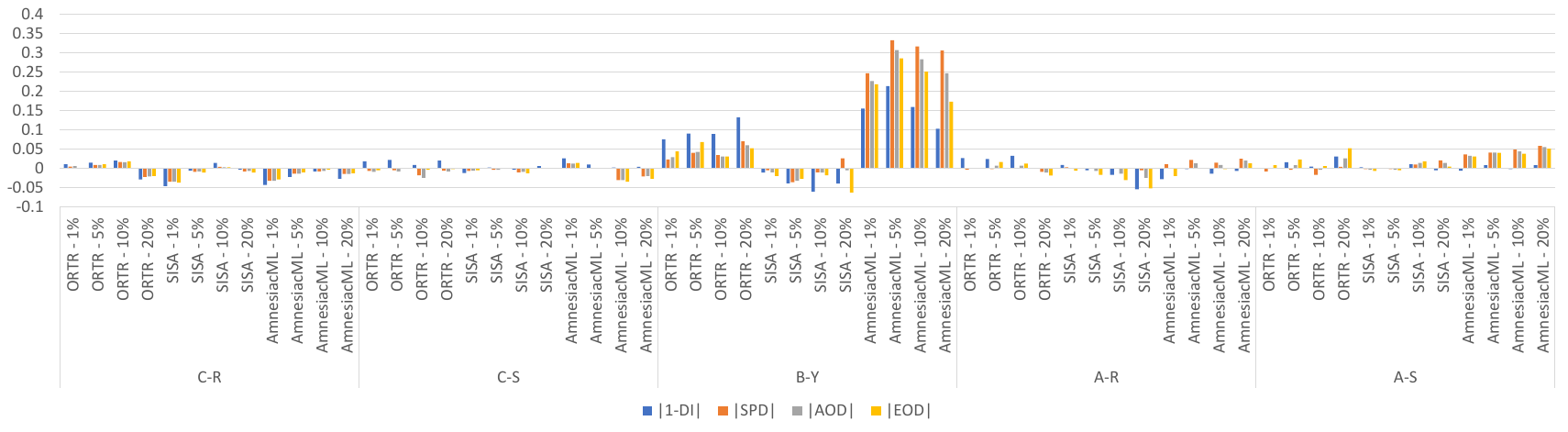}
  \caption{Difference of fairness between before and after the deletion. Value 0 indicates no fairness change, while positive values and negative values indicate worsen fairness and improved fairness respectively. 
  }
  \label{fig:rq2-fairness-diff}
\end{figure*}

\begin{flushleft}

\noindent \textbf{RQ1: (Initial training) What are the impacts of machine unlearning methods on fairness before the ``\textit{right to be forgotten}'' requests arrive?}
\end{flushleft}

The exact machine unlearning methods, such as SISA, modify how data is fed into machine learning models, affecting the fairness of these models before RTBF is requested, i.e., initial training. 
This research question is aimed to understand the impact of
machine unlearning methods on fairness in initial training. Specifically, we compare SISA with ORTR, a na\"ive approach built based on a MLP model. Note that the approximate machine unlearning methods, such as AmnesiacML, only update the ML models' parameters without modifying their architecture. We therefore ignore AmnesiacML in this research question. 

We evaluate the impact of SISA and ORTR on fairness across different numbers of shards (5, 10, 15, 20) and numbers of slices (1, 5). We execute the experiments on three different datasets, such as Adult, Bank, and COMPAS. For ease of observation, we denote Adult, Bank, and COMPAS as \textit{A}, \textit{B}, and \textit{C}, respectively. \textit{Sex}, \textit{Race}, and \textit{Age}, which are the sensitive features, are represented as \textit{S}, \textit{R}, and \textit{Y}, respectively.
The number of shards and the number of slices are represented as \textit{h} and \textit{c}, respectively. 
For example, given 1,000 instances, \textit{5h5c} means these instances are split into five shards. Each shard is then further split into five slices. In the end, each shard contains 200 instances and each slice includes 40 instances.

Figure~\ref{fig:rq1-fairness-1slice} shows the fairness evaluation results of SISA initial training with 5/10/15/20 shards and one slice. The baseline is ORTR. We can see that for some datasets and features, the $|1-\textrm{DI}|$ value gets better when the number of shards increases, including \textit{B-Y} and \textit{A-R}, while for \textit{C-S} and \textit{A-S} the value gets worse when the number of shards increases. Similarly, for other metrics, the trends are not always one way along the increasing number of shards. Although there are some tendencies within each dataset, overall across all datasets, we cannot come up with an interpretation towards any outstanding fairness impact from the SISA method and its number of shards.

In terms of performance, there is degradation of less than 10\% in accuracy for Adult and Bank datasets, while there is no apparent degradation for the COMPAS dataset. This could be because the COMPAS dataset is much smaller than the Adult and Bank datasets and has fewer useful features, making it easier to converge and less likely to experience performance degradation from data partitioning.

The fairness evaluation results of SISA at initial training with five slices are shown in Figure~\ref{fig:rq1-fairness-5slice}. Comparing the fairness between one slice and five slices, we find no noticeable difference across all fairness metrics. We have the similar observation on performance metrics shown in Figure~\ref{fig:rq1-performance-1slice} and Figure~\ref{fig:rq1-performance-5slice}, and this is expectedly identical to what was reported in the SISA paper \cite{sisa}.

\begin{tcolorbox}During initial training, no significant fairness impacts are observed from using machine unlearning methods, such as SISA. In addition, compared with ORTR, SISA has performance degradation on larger datasets.
\end{tcolorbox}

\begin{flushleft}
\textbf{RQ2: (Uniform distribution) What are the impacts of machine unlearning methods on fairness when the deleted data has uniform distribution?}
\end{flushleft}

A uniform data deletion strategy assumes that every instance has an equal possibility of being removed from trained models. In this research question, we want to explore how much these machine unlearning methods impact fairness when the deleted data is in uniform distribution.

For this research question, we employ a range of deletion rates from small to large (1\%, 5\%, 10\%, 20\%) chosen from the statistics~\cite{rtbf5years}. 
For SISA, we apply its default setting (i.e., \textit{5h1c}).
For AmnesiacML, we train its model according to the requirements in the paper~\cite{amnesiac}.

Figure~\ref{fig:rq2-bar-chart} presents the results of fairness in various deletion proportions. We see that there is no clear trend indicating which methods achieve better results across all datasets (i.e., Adult, Bank, and COMPAS) and sensitive features (\textit{Sex}, \textit{Race}, and \textit{Age}). Figure~\ref{fig:rq2-fairness-diff} shows the difference in fairness before and after applying the deletion strategy. It indicates that AmnesiacML is likely to be prone to fairness loss caused by this deletion strategy, while SISA is the most robust. However, the difference in fairness between before and after data deletion is unclear in Adult and COMPAS datasets. The main reason is that the deleted data is in uniform distribution, i.e., each instance has an equal probability of being removed from trained models, leading to similar fairness results in this setting. We also see that all methods have a relatively large variation in fairness on the Bank dataset. The reason is that this dataset is highly imbalanced compared to other datasets, such as Adult and COMPAS. Specifically, among 45,211 instances, only 963 instances (2.13\%) are labeled as negative instances in the Bank dataset.

\begin{figure*}[htbp]
  \centering
  \includegraphics[width=1.0\textwidth]{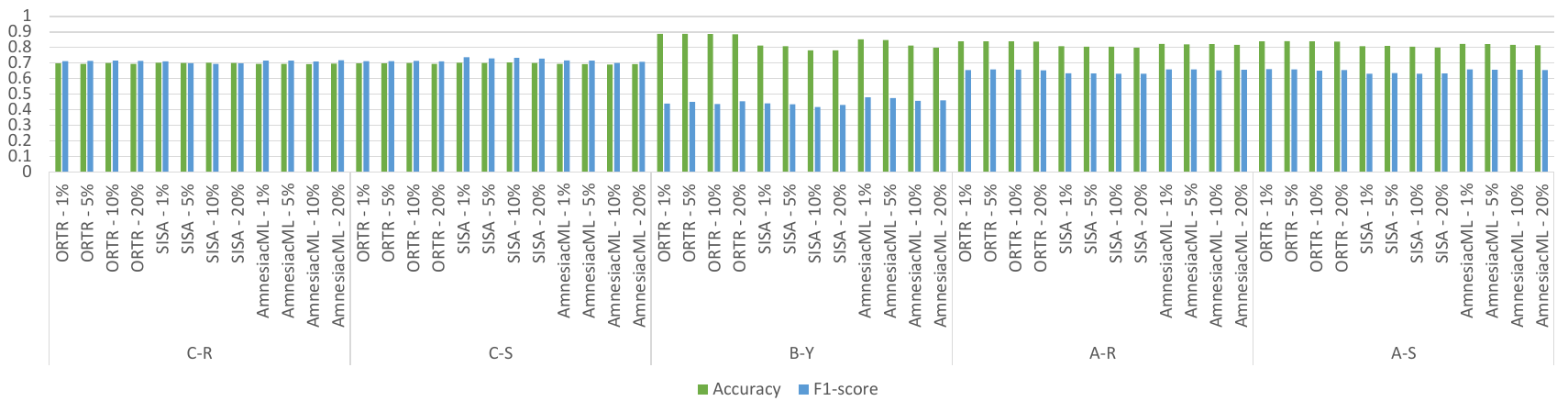}
  \caption{Performance (the higher, the better) results of different training methods after uniform data deletion under various deletion proportions.}
  \label{fig:rq2-peformance}
 \vspace{-6pt}
\end{figure*}

Figure~\ref{fig:rq2-peformance} illustrates the performance of this data deletion strategy. It shows that the deleted data has minimal impact in terms of performance on trained models. As the deletion proportion is 1\% - 20\%, we believe the deleted data might be insufficient to cause non-trivial performance degradation.

\begin{tcolorbox}
Under the data deletion of uniform distribution, the fairness is not clearly affected by machine unlearning methods, while ORTR outperforms SISA and AmnesiacML on performance metrics.
\end{tcolorbox}

\begin{flushleft}
\textbf{RQ3: (Non-uniform distribution) What are the impacts of machine unlearning methods on fairness when the deleted data has non-uniform distribution?}
\end{flushleft}

People from different groups have the equal right to send RTBF requests to remove their sensitive information, but they may have varied probabilities~\cite{eurobarometer}. In this research question, we aim to understand the impacts of machine unlearning methods on fairness when the deleted data has non-uniform distribution.

The simplest way to conduct the experiments is to remove the deleted data so that it has a similar distribution to the percentage of each group (privileged or unprivileged groups) for each sensitive feature on the whole dataset. As our datasets are imbalanced on some features, this RTBF simulation strategy highly likely leads to empty groups. To overcome this problem, we simplify our scenario by removing the data only from either the privileged group or the unprivileged group. Specifically, we remove 50\% of the data for each group, making the potential impact on fairness more apparent. Note that we assume the prior probability of a certain group (the privileged group or the unprivileged group) is known.

\begin{figure}[t!]
  \centering
  \begin{subfigure}[b]{0.48\textwidth}
  \centering
  \includegraphics[width=\textwidth]{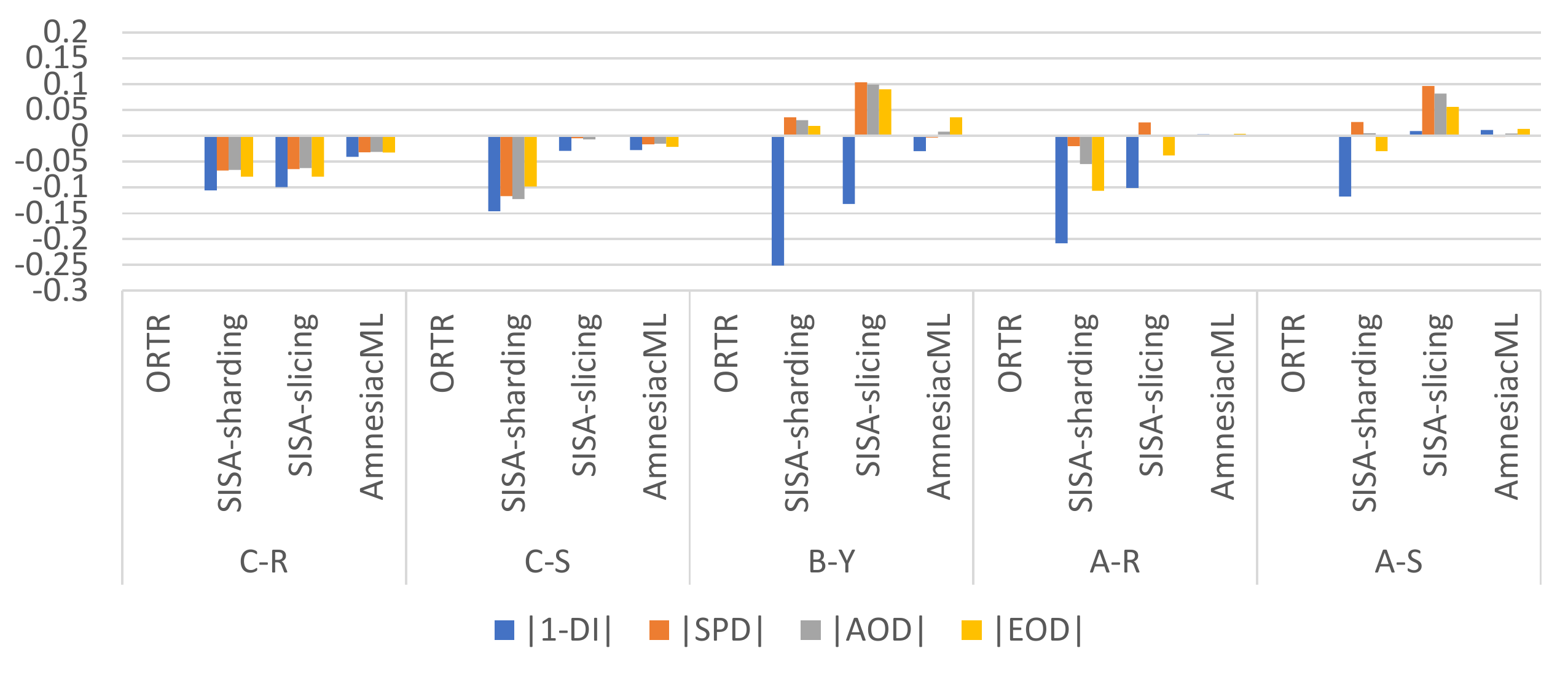}
  \caption{Fairness evaluation results after deletion from privileged group.}
  \label{fig:rq3-bar-chart-privileged}
  \end{subfigure}
  \begin{subfigure}[b]{0.48\textwidth}
  \centering
  \includegraphics[width=\textwidth]{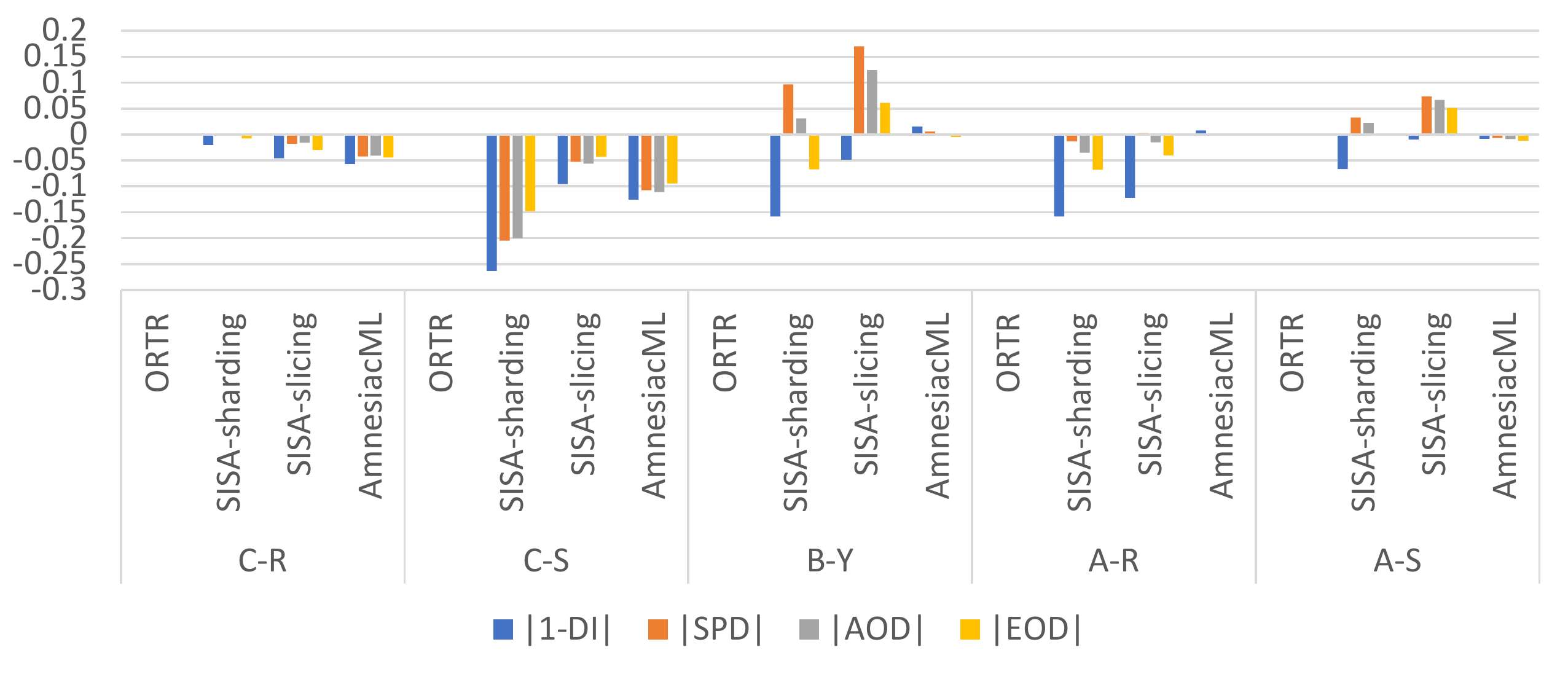}
  \caption{Fairness evaluation results after deletion from unprivileged group.}
  \label{fig:rq3-bar-chart-unprivileged}
  \end{subfigure}
  \caption{Fairness (the smaller, the better) evaluation results of non-uniform deletion. The results are shown as the distances from the ORTR results (baseline).}
  \label{fig:rq3-bar-chart}
 \vspace{-5pt}
\end{figure}

\begin{figure}[t!]
  \centering
  \begin{subfigure}[b]{0.24\textwidth}
  \centering
  \includegraphics[width=\textwidth]{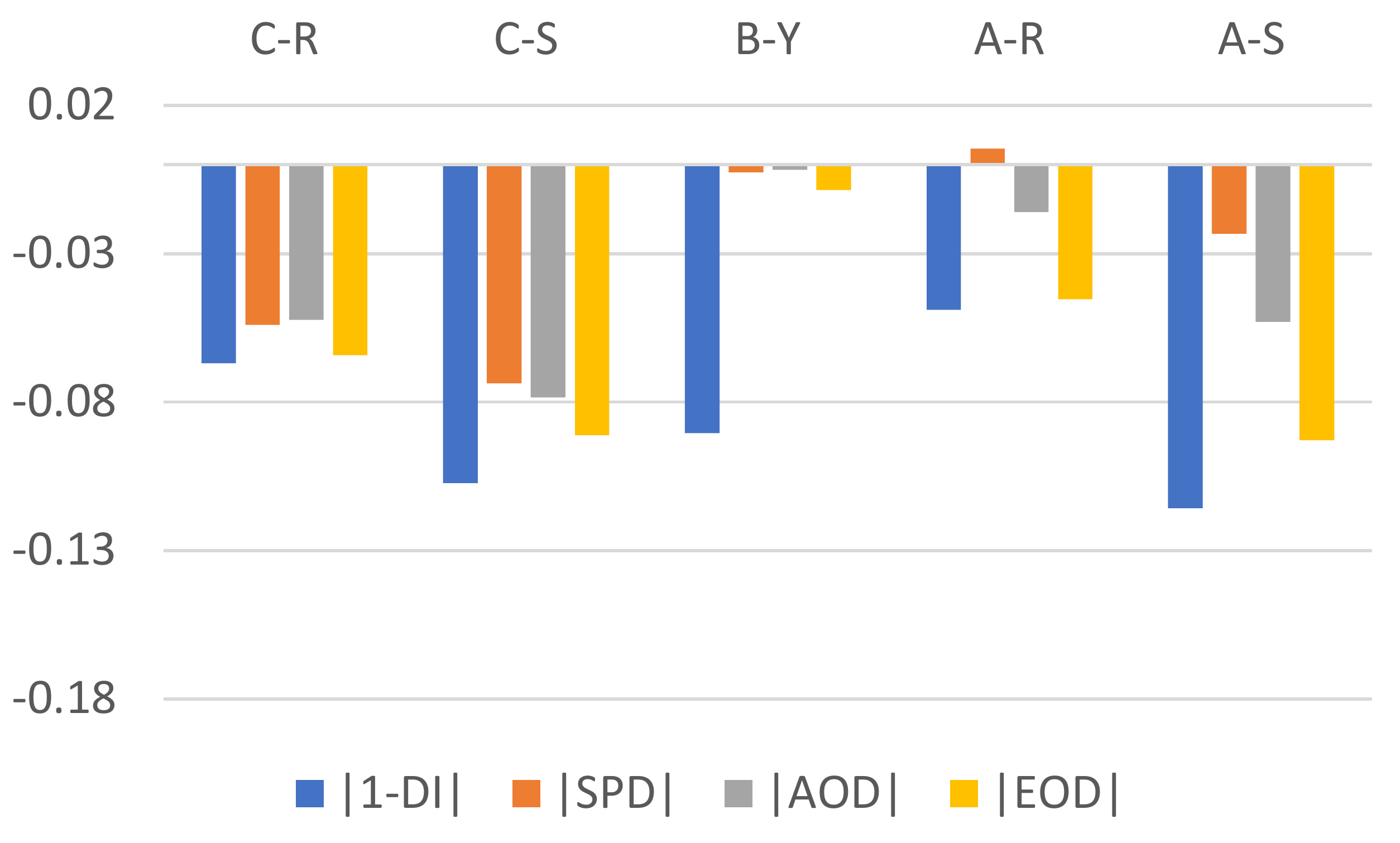}
  \caption{Fairness change after deletion using SISA with sharding strategy on privileged groups.}
  \label{fig:rq3-before-after-deletion-privileged}
  \end{subfigure}
  \begin{subfigure}[b]{0.24\textwidth}
  \centering
  \includegraphics[width=\textwidth]{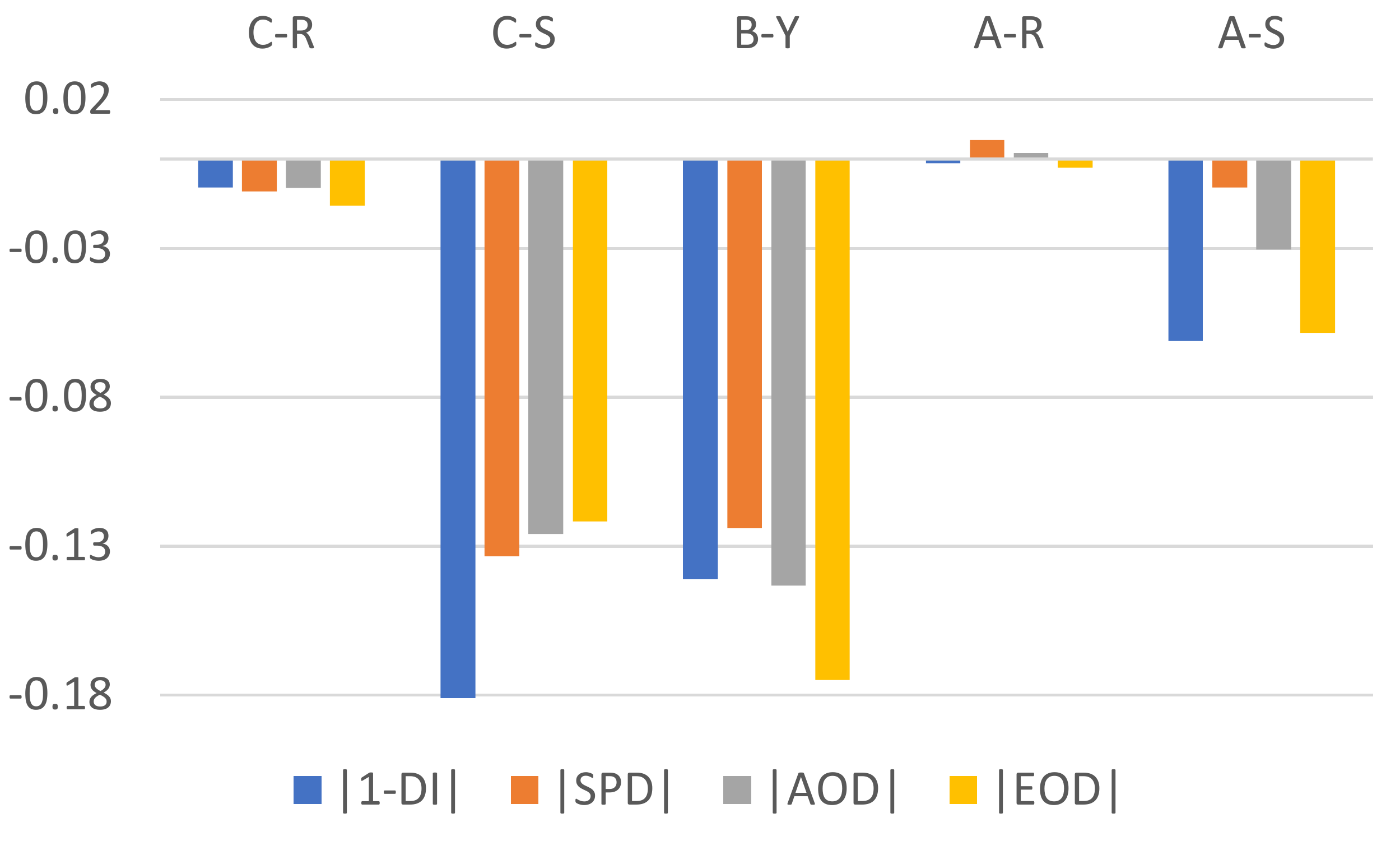}
  \caption{Fairness change after deletion using SISA with sharding strategy on unprivileged groups.}
  \label{fig:rq3-before-after-deletion-unprivileged}
  \end{subfigure}
  \caption{Fairness change after deletion using SISA with sharding strategy.}
  \label{fig:rq3-before-after-deletion-diff}
\end{figure}

\begin{figure}[htbp!]
  \centering
  \begin{subfigure}[b]{0.24\textwidth}
  \centering
  \includegraphics[width=\textwidth]{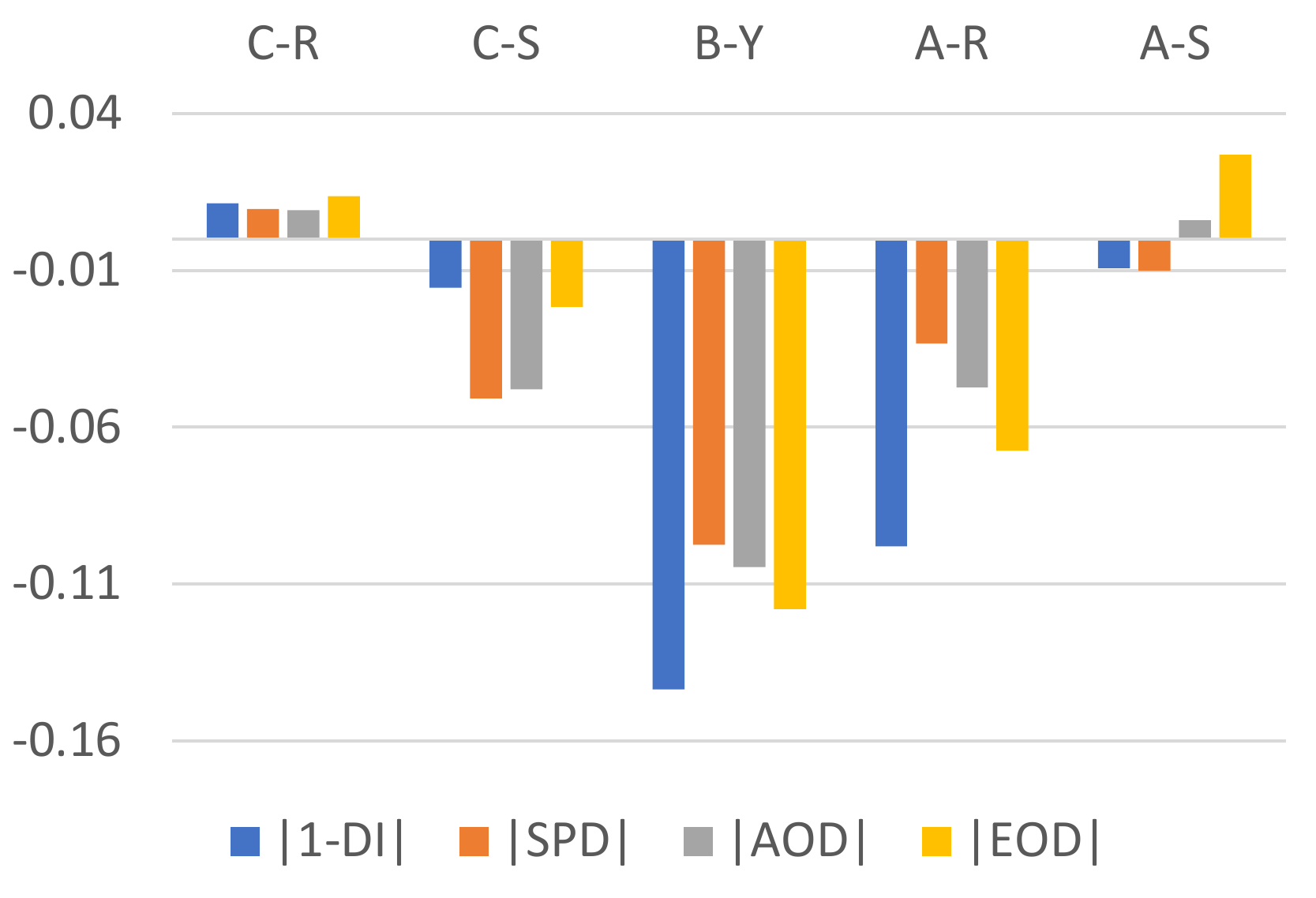}
  \caption{Fairness difference between SISA with and without applying sharding strategy on privileged groups.}
  \label{fig:rq3-with-without-sharding-strategy-diff-privileged}
  \end{subfigure}
  \begin{subfigure}[b]{0.24\textwidth}
  \centering
  \includegraphics[width=\textwidth]{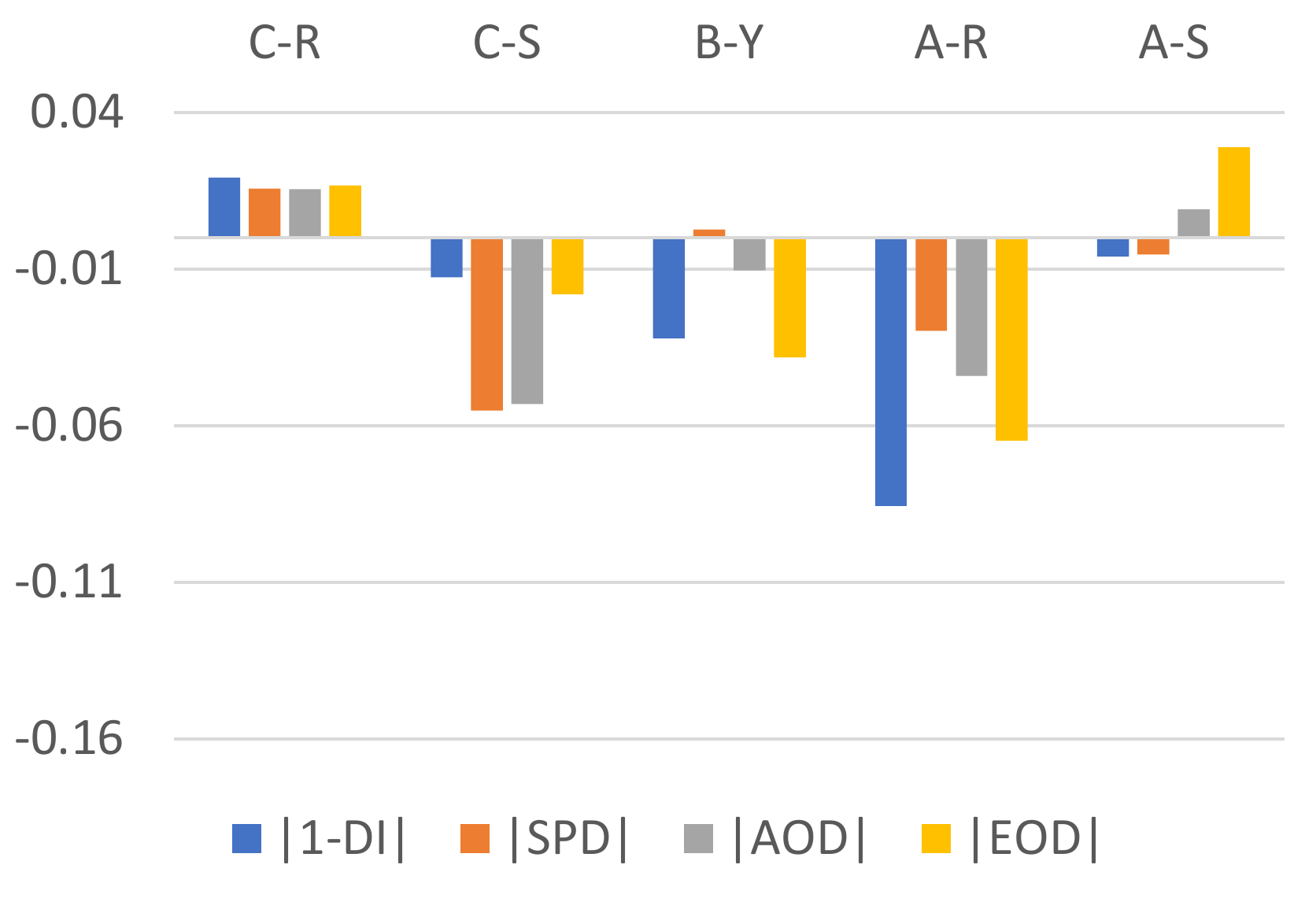}
  \caption{Fairness difference between SISA with and without applying sharding strategy on unprivileged groups.}
  \label{fig:rq3-with-without-sharding-strategy-diff-unprivileged}
  \end{subfigure}
  \caption{Fairness difference between SISA with and without applying sharding strategy.}
  \label{fig:rq3-with-without-sharding-strategy-diff}
\vspace{-5pt}
\end{figure}

Figure~\ref{fig:rq3-bar-chart-privileged} and Figure~\ref{fig:rq3-bar-chart-unprivileged} present the results on data deletion from the privileged group and the unprivileged group, respectively.
From the charts we can see that SISA with a sharding strategy (see Figure~\ref{fig:sisa-sharding}) achieve the best $|1-\textrm{DI}|$ values for nine out of ten combinations. 
Figure~\ref{fig:rq3-before-after-deletion-diff} shows that SISA with a sharding strategy may also have fairness improvements after data deletion. The extent of improvements varies under different datasets, such as Adult, Bank, and COMPAS, and different sensitive features, i.e., \textit{Sex}, \textit{Race}, and \textit{Age}. Furthermore, we plot the differences between SISA with and without the sharding strategy in Figure~\ref{fig:rq3-with-without-sharding-strategy-diff}. Overall, the fairness is likely to be improved across all metrics when the sharding strategy is applied. Moreover, such improvements are likely to happen on those datasets and sensitive features with more imbalanced distributions.

SISA with a slicing strategy (see Figure~\ref{fig:sisa-slicing}) is also likely to outperform ORTR on $|1-\textrm{DI}|$. However, it achieves less performance compared to SISA with a sharding strategy. For ORTR, we observe that the fairness changes between before and after retraining are weak. Similarly, AmnesiacML tends to be close to the ORTR across all indicators. 

Performance-wise, we observed no significant performance difference between before and after the data deletion, or between methods with and without strategies applied. The changes in performance indicators are always less than 5\%. The performance differences between different methods are likely to be inherited from the methods instead of escalated from unlearning strategies or distribution settings.

\begin{tcolorbox}
Under the data deletion of non-uniform distribution, SISA with a sharding strategy achieves better fairness. The performance has no significant degradation from deletion using machine unlearning methods.
\end{tcolorbox}

%% file: discussion/discussion.tex
\section{Discussion}
\label{sec:discussion}

Our research explores machine unlearning methods on fairness and has gained empirical observations about fairness regarding initial training, data deletion with uniform distribution, and data deletion with non-uniform distribution. We will discuss these observations as follows. 

Before the \textit{``right to be forgotten''} requests arrive, we see that there are no significant impacts of machine unlearning methods, such as SISA, on fairness. The observations also indicate that SISA achieves lower performance on large datasets, such as Adult and Bank, in this setting.

When the deleted data is in uniform distribution, there is no clear impact of machine unlearning methods on fairness. The observations also show that ORTR, a naive approach that retrains a model from scratch, outperforms SISA and AmnesiacML in terms of accuracy and F1 score on large datasets, i.e., Adult and Bank.

When the deleted data is in a non-uniform distribution, SISA with a sharding strategy (see Figure~\ref{fig:sisa-sharding}) is more likely to achieve better fairness compared to other models. Moreover, we also see that there is no significant performance difference between before and after the data deletion in machine unlearning methods.

%% file: threat-to-validity/threat-to-validity.tex
\section{Threat to Validity}
\label{sec:threat}
\subsection{Internal validity}
To perform our empirical study, we employed two machine unlearning methods on three AI fairness datasets. For machine unlearning algorithms, we reused existing implementations by following their open-source repositories. All three datasets are well-known and widely used by AI fairness researchers. We employed the AIF360 library to preprocess the datasets for fairness evaluation. We have carefully checked the code and data, but there might be some remaining errors. Although there was some randomness involved in the experiments, we have tried to minimize this threat by conducting experiments multiple times (5-fold cross-validation).

\subsection{External validity}
Threats to external validity refer to the generalizability of the study. In our experiments, we only used three AI fairness datasets, collected from three tasks, i.e., income prediction, customer churn prediction, and criminal detection, with a total of five protected classes and two machine unlearning methods to perform our experiments. We also performed two data deletion strategies. 
This may be a threat to external validity as these datasets, tasks, methods, and data deletion strategies may not be generalized beyond our studies. As the datasets and methods are widely adopted in AI fairness and machine unlearning research fields respectively, we believe that there is minimal threat to external validity. In the future, we plan to
investigate more machine unlearning methods and AI fairness datasets.

\subsection{Construct validity}
Threats to construct validity indicate evaluation metrics failing to be selective. We employed different evaluation metrics, widely used to measure fairness in machine learning models, to minimize threats to construct validity.

%% file: related-work/related-work.tex
\section{Related Work}
\label{sec:related_work}
This section introduces the work related to machine unlearning and AI fairness.

\subsection{Machine Unlearning}

Machine unlearning was first presented by Cao and Yang~\cite{towards-unlearning}. Its objective is to build a system that can remove the impact of a data point in the training data. Early works on \textit{machine unlearning} focused on traditional machine learning (ML) models, i.e., support vector machines, linear classification, logistic regression, etc., by facilitating incremental and decremental learning techniques to efficiently retrain ML models after adding or removing multiple data points from the training set~\cite{multi-incre-decre, regression-incre-decre, incre-decre-svm, incre-decre-linear}. Since then, machine unlearning has been extensively studied to reduce the computational cost of retraining deep learning (DL) models~\cite{towards-unlearning, baumhauer2022machine, sisa, golatkar2020forgetting, lifelong, amnesiac}. Specifically, there are two main research approaches for employing machine unlearning in deep neural networks, i.e., \textit{exact machine unlearning} and \textit{approximate machine unlearning}. 

The exact machine unlearning approach requires a new model to be trained from scratch by removing the deleted data from the training set. This approach ensures that the deleted data has no impact on the new model as we exclude it from the training set.
To make the retraining process more efficient, previous works~\cite{baumhauer2022machine, sisa} divided the training data into multiple disjoint shards and DL models were trained on each of these shards. Hence, when a request to remove data points from the training set comes, we only need to retrain the models containing the removed data points. The exact machine unlearning approach necessitates changes in the DL architecture, making testing and maintaining the DL system challenging. 

The approximate machine unlearning approach starts with the trained DL model and attempts to update its weights so that the model will no longer be affected by the removed data points from the training data. 
Izzo et al.~\cite{approximate} showed that we may achieve a linear time complexity in machine unlearning by updating a projective residual of the trained DL models. 
Guo et al.~\cite{certified-removal} and Golatkar et al.~\cite{selective-forgetting} employed a Newton step on the model weights to eliminate the influence of removed data points. 
Graves et al.~\cite{amnesiac} later proposed an amnesiac unlearning method by storing a list of batches and their updated weights; hence, DL models only need to undo the updated weights from the batches containing the removed data points. 
The approximate machine unlearning approach is more computationally efficient than the exact machine learning approach. However, we are unsure whether the removed data points have been completely forgotten in the trained model. 

Although machine unlearning methods have been comprehensively studied, their fairness has not been investigated in the process of building machine unlearning systems. To fill in this gap, we perform an extensive study on the two machine unlearning approaches, i.e., extract and approximate, to reveal their fairness implications.

\subsection{AI Fairness}

\textit{AI fairness} or machine learning (ML) fairness has been deeply investigated during the last decade~\cite{fairness-re, RE-AI, fairness-survey, fairness-testing, zhang2021ignorance, biswas2020machine, dwork2012fairness}. Its basic idea is that the prediction model should not be biased between different individuals or groups from the protected attribute class (e.g., race, sex, familial status, etc.). There are mainly two major types of AI fairness, i.e., \textit{group fairness} and \textit{individual fairness}~\cite{fairness-survey, dwork2012fairness}. 

Group fairness requires the prediction model to produce different predictive results for different groups in the protected attribute class. Several studies proposed a specific kind of utility maximization decision function to satisfy a fairness constraint and derive optimal fairness decisions~\cite{corbett2017algorithmic, hardt2016equality, menon2018cost, baumann2022enforcing}. Hardt et al.~\cite{hardt2016equality} employed the Bayes optimal non-discriminant to derive fairness in a classification model. Corbett-Davies et al.~\cite{corbett2017algorithmic} considered AI fairness as a constrained optimization problem to maximize accuracy while satisfying group fairness constraints. Menon and Williamson~\cite{menon2018cost} investigated the trade-off between accuracy and group fairness in AI models and proposed a threshold function for the fairness problem. Group fairness often ignores the individual characteristics of the group, leading to permit unfairness in training ML models~\cite{kearns2018preventing}. 

Individual fairness on the other hand expects the prediction model to produce similar predictive results among similar individuals who are only different in protected attributes. 
Udeshi et al.~\cite{udeshi2018automated} presented Aequitas, a fully automated and directed test generation framework, to generate test inputs and improve the individual fairness of ML models. Aggarwal et al.~\cite{aggarwal2019black} employed both symbolic execution (together with local explainability) to identify factors making decisions and then generate test inputs. Sun et al.~\cite{sun2020automatic} combined both input mutation and metamorphic relations to improve the fairness of machine translation.

Other works explore the effectiveness and efficiency of existing ML methods for software fairness~\cite{biswas2020machine, chakraborty2020fairway, zhang2021ignorance}. Specifically, researchers focus on improving fairness in ML systems by leveraging mitigation techniques~\cite{biswas2020machine}, removing biased instances in training data~\cite{chakraborty2020fairway}, or improving the quality of features in the datasets~\cite{zhang2021ignorance}.

Even though AI fairness has been widely adopted, its properties have not been revealed in machine unlearning. We perform an empirical study on the three AI fairness datasets, i.e., Adult, Bank, and COMPAS to understand the impacts of machine unlearning models on fairness.

%% file: conclusion/conclusion.tex
\section{Conclusion and Future Work}
\label{sec:conclusion}
Machine unlearning emerges with the need to implement the \textit{``right to be forgotten''} (RTBF) efficiently while existing studies overlook its impact on fairness. To the best of our knowledge, we are the first to perform an empirical study on the impacts of machine unlearning methods on fairness. We designed and conducted experiments on two typical machine unlearning methods (SISA and AmnesiacML) along with a retraining method (ORTR) using three fairness datasets under three different deletion strategies. We found that SISA leads to better fairness compared with AmnesiacML and ORTR, while initial training and uniform data deletion do not necessarily affect the fairness of all three methods. Our research has shed light on fairness implications of machine unlearning and provided knowledge for software engineers about the trade-offs when considering machine unlearning methods as a solution for RTBF. In the future, more research efforts are needed to broaden the understanding of fairness implications into other machine unlearning methods as well as investigate the underlying causes of their impact on fairness.

%% file: main.bbl
% Generated by IEEEtran.bst, version: 1.14 (2015/08/26)

%% file: main.bbl
\begin{thebibliography}{10}
\providecommand{\url}[1]{#1}
\csname url@samestyle\endcsname
\providecommand{\newblock}{\relax}
\providecommand{\bibinfo}[2]{#2}
\providecommand{\BIBentrySTDinterwordspacing}{\spaceskip=0pt\relax}
\providecommand{\BIBentryALTinterwordstretchfactor}{4}
\providecommand{\BIBentryALTinterwordspacing}{\spaceskip=\fontdimen2\font plus
\BIBentryALTinterwordstretchfactor\fontdimen3\font minus
  \fontdimen4\font\relax}
\providecommand{\BIBforeignlanguage}[2]{{%
\expandafter\ifx\csname l@#1\endcsname\relax
\typeout{** WARNING: IEEEtran.bst: No hyphenation pattern has been}%
\typeout{** loaded for the language `#1'. Using the pattern for}%
\typeout{** the default language instead.}%
\else
\language=\csname l@#1\endcsname
\fi
#2}}
\providecommand{\BIBdecl}{\relax}
\BIBdecl

\bibitem{singhal2022survey}
P.~Singhal, P.~K. Srivastava, A.~K. Tiwari, and R.~K. Shukla, ``A survey:
  Approaches to facial detection and recognition with machine learning
  techniques,'' in \emph{Proceedings of Second Doctoral Symposium on
  Computational Intelligence}.\hskip 1em plus 0.5em minus 0.4em\relax Springer,
  2022, pp. 103--125.

\bibitem{li2011design}
W.~Li, J.~Matejka, T.~Grossman, J.~A. Konstan, and G.~Fitzmaurice, ``Design and
  evaluation of a command recommendation system for software applications,''
  \emph{ACM Transactions on Computer-Human Interaction (TOCHI)}, vol.~18,
  no.~2, pp. 1--35, 2011.

\bibitem{rudin2018optimized}
C.~Rudin and B.~Ustun, ``Optimized scoring systems: Toward trust in machine
  learning for healthcare and criminal justice,'' \emph{Interfaces}, vol.~48,
  no.~5, pp. 449--466, 2018.

\bibitem{ratner2019role}
A.~J. Ratner, B.~Hancock, and C.~R{\'e}, ``The role of massively multi-task and
  weak supervision in software 2.0.'' in \emph{CIDR}, 2019.

\bibitem{gdpr}
``General data protection regulation,'' \url{https://gdpr-info.eu/}.

\bibitem{ccpa}
``California consumer privacy act,'' \url{https://oag.ca.gov/privacy/ccpa}.

\bibitem{PIPEDA}
``Announcement: Privacy commissioner seeks federal court determination on key
  issue for canadians’ online reputation,''
  \url{https://www.priv.gc.ca/en/opc-news/news-and-announcements/2018/an_181010/}.

\bibitem{rtbf}
J.~Rosen, ``The right to be forgotten,'' \emph{Stan. L. Rev. Online}, vol.~64,
  p.~88, 2011.

\bibitem{responsible-data-management}
J.~Stoyanovich, B.~Howe, and H.~V. Jagadish, ``{Responsible data management},''
  \emph{Proceedings of the VLDB Endowment}, vol.~13, no.~12, pp. 3474--3488,
  2020.

\bibitem{thompson2020computational}
N.~C. Thompson, K.~Greenewald, K.~Lee, and G.~F. Manso, ``The computational
  limits of deep learning,'' \emph{arXiv preprint arXiv:2007.05558}, 2020.

\bibitem{towards-unlearning}
Y.~Cao and J.~Yang, ``Towards making systems forget with machine unlearning,''
  in \emph{2015 IEEE Symposium on Security and Privacy}.\hskip 1em plus 0.5em
  minus 0.4em\relax IEEE, 2015, pp. 463--480.

\bibitem{baumhauer2022machine}
T.~Baumhauer, P.~Sch{\"o}ttle, and M.~Zeppelzauer, ``Machine unlearning: Linear
  filtration for logit-based classifiers,'' \emph{Machine Learning}, pp. 1--24,
  2022.

\bibitem{sisa}
L.~Bourtoule, V.~Chandrasekaran, C.~A. Choquette-Choo, H.~Jia, A.~Travers,
  B.~Zhang, D.~Lie, and N.~Papernot, ``Machine unlearning,'' in \emph{2021 IEEE
  Symposium on Security and Privacy (SP)}.\hskip 1em plus 0.5em minus
  0.4em\relax IEEE, 2021, pp. 141--159.

\bibitem{golatkar2020forgetting}
A.~Golatkar, A.~Achille, and S.~Soatto, ``Forgetting outside the box: Scrubbing
  deep networks of information accessible from input-output observations,'' in
  \emph{European Conference on Computer Vision}.\hskip 1em plus 0.5em minus
  0.4em\relax Springer, 2020, pp. 383--398.

\bibitem{lifelong}
M.~Du, Z.~Chen, C.~Liu, R.~Oak, and D.~Song, ``Lifelong anomaly detection
  through unlearning,'' in \emph{Proceedings of the 2019 ACM SIGSAC Conference
  on Computer and Communications Security}, 2019, pp. 1283--1297.

\bibitem{amnesiac}
\BIBentryALTinterwordspacing
L.~Graves, V.~Nagisetty, and V.~Ganesh, ``Amnesiac machine learning,''
  \emph{Proceedings of the AAAI Conference on Artificial Intelligence},
  vol.~35, no.~13, pp. 11\,516--11\,524, May 2021. [Online]. Available:
  \url{https://ojs.aaai.org/index.php/AAAI/article/view/17371}
\BIBentrySTDinterwordspacing

\bibitem{fairness-re}
A.~Finkelstein, M.~Harman, S.~A. Mansouri, J.~Ren, and Y.~Zhang, ``“fairness
  analysis” in requirements assignments,'' in \emph{2008 16th IEEE
  International Requirements Engineering Conference}.\hskip 1em plus 0.5em
  minus 0.4em\relax IEEE, 2008, pp. 115--124.

\bibitem{RE-AI}
K.~Ahmad, M.~Bano, M.~Abdelrazek, C.~Arora, and J.~Grundy, ``What’s up with
  requirements engineering for artificial intelligence systems?'' in \emph{2021
  IEEE 29th International Requirements Engineering Conference (RE)}.\hskip 1em
  plus 0.5em minus 0.4em\relax IEEE, 2021, pp. 1--12.

\bibitem{fairness-survey}
N.~Mehrabi, F.~Morstatter, N.~Saxena, K.~Lerman, and A.~Galstyan, ``A survey on
  bias and fairness in machine learning,'' \emph{ACM Computing Surveys (CSUR)},
  vol.~54, no.~6, pp. 1--35, 2021.

\bibitem{fairness-testing}
J.~M. Zhang, M.~Harman, L.~Ma, and Y.~Liu, ``{Machine Learning Testing: Survey,
  Landscapes and Horizons},'' \emph{IEEE Transactions on Software Engineering},
  vol.~48, no.~1, pp. 1--36, 2020.

\bibitem{zhang2021ignorance}
J.~M. Zhang and M.~Harman, ``"ignorance and prejudice" in software fairness,''
  in \emph{2021 IEEE/ACM 43rd International Conference on Software Engineering
  (ICSE)}.\hskip 1em plus 0.5em minus 0.4em\relax IEEE, 2021, pp. 1436--1447.

\bibitem{biswas2020machine}
S.~Biswas and H.~Rajan, ``Do the machine learning models on a crowd sourced
  platform exhibit bias? an empirical study on model fairness,'' in
  \emph{Proceedings of the 28th ACM joint meeting on European software
  engineering conference and symposium on the foundations of software
  engineering}, 2020, pp. 642--653.

\bibitem{dwork2012fairness}
C.~Dwork, M.~Hardt, T.~Pitassi, O.~Reingold, and R.~Zemel, ``Fairness through
  awareness,'' in \emph{Proceedings of the 3rd innovations in theoretical
  computer science conference}, 2012, pp. 214--226.

\bibitem{chakraborty2020fairway}
J.~Chakraborty, S.~Majumder, Z.~Yu, and T.~Menzies, ``Fairway: A way to build
  fair ml software,'' in \emph{Proceedings of the 28th ACM Joint Meeting on
  European Software Engineering Conference and Symposium on the Foundations of
  Software Engineering}, 2020, pp. 654--665.

\bibitem{murakawa2014first}
N.~Murakawa, \emph{The first civil right: How liberals built prison
  America}.\hskip 1em plus 0.5em minus 0.4em\relax Oxford University Press,
  2014.

\bibitem{leung2007naive}
K.~M. Leung, ``Naive bayesian classifier,'' \emph{Polytechnic University
  Department of Computer Science/Finance and Risk Engineering}, vol. 2007, pp.
  123--156, 2007.

\bibitem{upton2001strategic}
N.~Upton, E.~J. Teal, and J.~T. Felan, ``Strategic and business planning
  practices of fast growth family firms,'' \emph{Journal of small business
  management}, vol.~39, no.~1, pp. 60--72, 2001.

\bibitem{eurobarometer}
``Charter of fundamental rights and general data protection regulation,''
  \url{https://europa.eu/eurobarometer/surveys/detail/2222}, 2019.

\bibitem{recommendation}
C.~Chen, F.~Sun, M.~Zhang, and B.~Ding, ``Recommendation unlearning,'' in
  \emph{Proceedings of the ACM Web Conference 2022}, 2022, pp. 2768--2777.

\bibitem{coded}
N.~Aldaghri, H.~Mahdavifar, and A.~Beirami, ``Coded machine unlearning,''
  \emph{IEEE Access}, vol.~9, pp. 88\,137--88\,150, 2021.

\bibitem{graph-eraser}
M.~Chen, Z.~Zhang, T.~Wang, M.~Backes, M.~Humbert, and Y.~Zhang, ``Graph
  unlearning,'' \emph{arXiv preprint arXiv:2103.14991}, 2021.

\bibitem{di}
A.~Chouldechova, ``Fair prediction with disparate impact: A study of bias in
  recidivism prediction instruments,'' \emph{Big data}, vol.~5, no.~2, pp.
  153--163, 2017.

\bibitem{spd}
T.~Calders and S.~Verwer, ``Three naive bayes approaches for
  discrimination-free classification,'' \emph{Data mining and knowledge
  discovery}, vol.~21, no.~2, pp. 277--292, 2010.

\bibitem{equal-op}
M.~Hardt, E.~Price, and N.~Srebro, ``Equality of opportunity in supervised
  learning,'' \emph{Advances in neural information processing systems},
  vol.~29, 2016.

\bibitem{aggarwal2019black}
A.~Aggarwal, P.~Lohia, S.~Nagar, K.~Dey, and D.~Saha, ``Black box fairness
  testing of machine learning models,'' in \emph{Proceedings of the 2019 27th
  ACM Joint Meeting on European Software Engineering Conference and Symposium
  on the Foundations of Software Engineering}, 2019, pp. 625--635.

\bibitem{heart}
``Heart disease data set,''
  \url{https://archive.ics.uci.edu/ml/datasets/heart+disease}, 1988.

\bibitem{compas}
\url{https://github.com/propublica/compas-analysis}, 2017.

\bibitem{aif360}
R.~K. Bellamy, K.~Dey, M.~Hind, S.~C. Hoffman, S.~Houde, K.~Kannan, P.~Lohia,
  J.~Martino, S.~Mehta, A.~Mojsilovi{\'c} \emph{et~al.}, ``Ai fairness 360: An
  extensible toolkit for detecting and mitigating algorithmic bias,'' \emph{IBM
  Journal of Research and Development}, vol.~63, no. 4/5, pp. 4--1, 2019.

\bibitem{ramchoun2016multilayer}
H.~Ramchoun, Y.~Ghanou, M.~Ettaouil, and M.~A. Janati~Idrissi, ``Multilayer
  perceptron: Architecture optimization and training,'' \emph{International
  Journal of Interactive Multimedia and Artificial Intelligence}, 2016.

\bibitem{martinez2018taming}
M.~Martinez and R.~Stiefelhagen, ``Taming the cross entropy loss,'' in
  \emph{German Conference on Pattern Recognition}.\hskip 1em plus 0.5em minus
  0.4em\relax Springer, 2018, pp. 628--637.

\bibitem{adult}
``Adult data set,'' \url{https://archive.ics.uci.edu/ml/datasets/adult}, 1996.

\bibitem{bank}
``Bank marketing data set,''
  \url{https://archive.ics.uci.edu/ml/datasets/Bank+Marketing}, 2012.

\bibitem{rtbf5years}
T.~Bertram, E.~Bursztein, S.~Caro, H.~Chao, R.~C. Feman, P.~Fleischer,
  A.~Gustafsson, J.~Hemerly, C.~Hibbert, L.~Invernizzi, L.~K. Donnelly,
  J.~Ketover, J.~Laefer, P.~Nicholas, Y.~Niu, H.~Obhi, D.~Price, A.~Strait,
  K.~Thomas, and A.~Verney, ``Five years of the right to be forgotten,'' in
  \emph{Proceedings of the Conference on Computer and Communications Security},
  2019.

\bibitem{gunawardana2009survey}
A.~Gunawardana and G.~Shani, ``A survey of accuracy evaluation metrics of
  recommendation tasks.'' \emph{Journal of Machine Learning Research}, vol.~10,
  no.~12, 2009.

\bibitem{dalianis2018evaluation}
H.~Dalianis, ``Evaluation metrics and evaluation,'' in \emph{Clinical text
  mining}.\hskip 1em plus 0.5em minus 0.4em\relax Springer, 2018, pp. 45--53.

\bibitem{multi-incre-decre}
M.~Karasuyama and I.~Takeuchi, ``Multiple incremental decremental learning of
  support vector machines,'' \emph{IEEE Transactions on Neural Networks},
  vol.~21, no.~7, pp. 1048--1059, 2010.

\bibitem{regression-incre-decre}
C.-H. Tsai, C.-Y. Lin, and C.-J. Lin, ``Incremental and decremental training
  for linear classification,'' in \emph{Proceedings of the 20th ACM SIGKDD
  international conference on Knowledge discovery and data mining}, 2014, pp.
  343--352.

\bibitem{incre-decre-svm}
G.~Cauwenberghs and T.~Poggio, ``Incremental and decremental support vector
  machine learning,'' \emph{Advances in neural information processing systems},
  vol.~13, 2000.

\bibitem{incre-decre-linear}
E.~Romero, I.~Barrio, and L.~Belanche, ``Incremental and decremental learning
  for linear support vector machines,'' in \emph{Artificial Neural Networks --
  ICANN 2007}, J.~M. de~S{\'a}, L.~A. Alexandre, W.~Duch, and D.~Mandic,
  Eds.\hskip 1em plus 0.5em minus 0.4em\relax Berlin, Heidelberg: Springer
  Berlin Heidelberg, 2007, pp. 209--218.

\bibitem{approximate}
Z.~{Izzo}, M.~A. {Smart}, K.~{Chaudhuri}, and J.~{Zou}, ``{Approximate Data
  Deletion from Machine Learning Models},'' \emph{arXiv e-prints}, p.
  arXiv:2002.10077, Feb. 2020.

\bibitem{certified-removal}
\BIBentryALTinterwordspacing
C.~Guo, T.~Goldstein, A.~Hannun, and L.~Van Der~Maaten, ``Certified data
  removal from machine learning models,'' in \emph{Proceedings of the 37th
  International Conference on Machine Learning}, ser. Proceedings of Machine
  Learning Research, H.~D. III and A.~Singh, Eds., vol. 119.\hskip 1em plus
  0.5em minus 0.4em\relax PMLR, 13--18 Jul 2020, pp. 3832--3842. [Online].
  Available: \url{https://proceedings.mlr.press/v119/guo20c.html}
\BIBentrySTDinterwordspacing

\bibitem{selective-forgetting}
A.~Golatkar, A.~Achille, and S.~Soatto, ``Eternal sunshine of the spotless net:
  Selective forgetting in deep networks,'' in \emph{Proceedings of the IEEE/CVF
  Conference on Computer Vision and Pattern Recognition}, 2020, pp. 9304--9312.

\bibitem{corbett2017algorithmic}
S.~Corbett-Davies, E.~Pierson, A.~Feller, S.~Goel, and A.~Huq, ``Algorithmic
  decision making and the cost of fairness,'' in \emph{Proceedings of the 23rd
  acm sigkdd international conference on knowledge discovery and data mining},
  2017, pp. 797--806.

\bibitem{hardt2016equality}
M.~Hardt, E.~Price, and N.~Srebro, ``Equality of opportunity in supervised
  learning,'' \emph{Advances in neural information processing systems},
  vol.~29, 2016.

\bibitem{menon2018cost}
A.~K. Menon and R.~C. Williamson, ``The cost of fairness in binary
  classification,'' in \emph{Conference on Fairness, Accountability and
  Transparency}.\hskip 1em plus 0.5em minus 0.4em\relax PMLR, 2018, pp.
  107--118.

\bibitem{baumann2022enforcing}
J.~Baumann, A.~Hann{\'a}k, and C.~Heitz, ``Enforcing group fairness in
  algorithmic decision making: Utility maximization under sufficiency,''
  \emph{arXiv preprint arXiv:2206.02237}, 2022.

\bibitem{kearns2018preventing}
M.~Kearns, S.~Neel, A.~Roth, and Z.~S. Wu, ``Preventing fairness
  gerrymandering: Auditing and learning for subgroup fairness,'' in
  \emph{International Conference on Machine Learning}.\hskip 1em plus 0.5em
  minus 0.4em\relax PMLR, 2018, pp. 2564--2572.

\bibitem{udeshi2018automated}
S.~Udeshi, P.~Arora, and S.~Chattopadhyay, ``Automated directed fairness
  testing,'' in \emph{Proceedings of the 33rd ACM/IEEE International Conference
  on Automated Software Engineering}, 2018, pp. 98--108.

\bibitem{sun2020automatic}
Z.~Sun, J.~M. Zhang, M.~Harman, M.~Papadakis, and L.~Zhang, ``Automatic testing
  and improvement of machine translation,'' in \emph{Proceedings of the
  ACM/IEEE 42nd International Conference on Software Engineering}, 2020, pp.
  974--985.

\end{thebibliography}
